\documentclass[twocolumn,showpacs,preprintnumbers,nofootinbib,prd,
superscriptaddress,10pt]{revtex4-1}

\usepackage{amsmath,amssymb}
\usepackage[normalem]{ulem}
\usepackage{textcomp}
\usepackage{hyperref}
\usepackage{bm}
\usepackage{graphicx}
\usepackage{psfrag}
\usepackage[usenames,dvipsnames]{xcolor}
\usepackage[utf8]{inputenc}
\usepackage[english, capitalise]{cleveref}
\crefname{chapter}{Chap.}{Chap.}
\crefname{section}{Sec.}{Secs.}
\Crefname{chapter}{Chapter}{Chapters}
\Crefname{section}{Section}{Sections}
\Crefname{eqs}{Eqs.}{Eqs.}		
\Crefformat{eqs}{Eqs.~(#2#1#3)}	

\definecolor{darkgreen}{rgb}{0,0.5,0}
\hypersetup{
	bookmarks=true,         
	unicode=false,          
	pdftoolbar=true,        
	pdfmenubar=true,        
	pdffitwindow=false,     
	pdfstartview={FitH},    
	pdftitle={Gravitational-wave amplitudes for compact binaries in eccentric orbits at the third
		post-Newtonian order: Memory contributions},
	pdfauthor={M.~Ebersold, Y.~Boetzel, G.~Faye, C.~K.~Mishra, B.~Iyer, P.~Jetzer},
	pdfsubject={Gravitational Waves},
	pdfkeywords={gravitational waves, eccentric compact binaries, general relativity},
	pdfnewwindow=true,      
	colorlinks=true,       
	linkcolor=red,          
	citecolor=cyan,        
	filecolor=magenta,      
	urlcolor=darkgreen,           
	linktocpage=true
}

\allowdisplaybreaks[4]

\newcommand{\bigO}{\mathcal{O}}

\def\no{\nonumber\\}
\newcommand{\eg}{\gamma_\textnormal{E}}
\newcommand{\inst}{\textnormal{inst}}
\newcommand{\hered}{\textnormal{hered}}
\newcommand{\tail}{\textnormal{tail}}
\newcommand{\postad}{\textnormal{post-ad}}
\newcommand{\mem}{\textnormal{mem}}
\newcommand{\memdc}{\textnormal{DC}}
\newcommand{\memosc}{\textnormal{osc}}
\newcommand{\ud}{\mathrm{d}}
\newcommand{\ui}{\mathrm{i}} 
\newcommand{\ue}{\mathrm{e}} 

\newcommand{\GW}{\textnormal{GW}}
\newcommand{\newt}{\textnormal{Newt}}
\newcommand{\lead}{\textnormal{Lead}}
\newcommand{\pn}[1]{\textnormal{{#1}PN}}

\newcommand{\xb}{\bar{x}}

\newcommand{\eb}{\bar{e}}

\graphicspath{{./figures/}}

\begin{document}

\title{Gravitational-wave amplitudes for compact binaries in eccentric orbits at the third 
	post-Newtonian order: Memory contributions}

\date{\today}

\author{Michael Ebersold}
\affiliation{Physik-Institut, Universit\"at Z\"urich, Winterthurerstrasse 190, 8057 Z\"urich}

\author{Yannick Boetzel}
\affiliation{Physik-Institut, Universit\"at Z\"urich, Winterthurerstrasse 190, 8057 Z\"urich}

\author{Guillaume Faye}
\affiliation{$\mathcal{G}\mathbb{R}\varepsilon{\mathbb{C}}\mathcal{O}$, 
	Institut d'Astrophysique de Paris, UMR 7095, CNRS, Sorbonne Universit{\'e}, 
	98\textsuperscript{bis} boulevard Arago, 75014 Paris, France}

\author{Chandra Kant Mishra}
\affiliation{Department of Physics, Indian Institute of Technology, Madras, Chennai 600036, India}

\author{Bala R. Iyer}
\affiliation{International Centre for Theoretical Sciences, Tata Institute of Fundamental Research, 
	Bangalore 560089, India}

\author{Philippe Jetzer}
\affiliation{Physik-Institut, Universit\"at Z\"urich, Winterthurerstrasse 190, 8057 Z\"urich}

\begin{abstract}
We compute the \emph{nonlinear memory} contributions to the gravitational-wave amplitudes for
compact binaries in eccentric orbits at the third post-Newtonian (3PN) order in general relativity.
These contributions are \emph{hereditary} in nature as they are sourced by gravitational waves
emitted during the binary's entire dynamical past. Combining these with already available
instantaneous and tail contributions we get the complete 3PN accurate gravitational waveform.
\end{abstract}


\maketitle

\section{Introduction}\label{sec: introduction}

The observation of the first gravitational-wave (GW) signal by LIGO and Virgo opened up the new
field of gravitational-wave astronomy~\cite{LIGO, Virgo, GEO600, GW150914}. So far, ten confirmed
binary black hole mergers and one binary neutron star coalescence have been reported~\cite{GW151226,
GW170104, GW170814, GW170817, GW170608, GWTC-1}. KAGRA~\cite{KAGRA} is expected to join the global
network of detectors later this year, followed by LIGO-India in 2025~\cite{LIGO-India}, leading to
improved parameter estimation and source localization. These ground-based detectors are sensitive
to the decahertz--kilohertz frequency of the GW spectrum. In the future the space-based detector
LISA~\cite{amaro-2017} will probe lower frequencies (around the millihertz range), and pulsar
timing arrays (PTAs) may measure ultralow (nanohertz) frequency GWs~\cite{IPTA-2013}.

Currently (and this will most probably not change in the future), compact binaries are the most
important sources of observable GW signals. The events detected so far have all been found using
circular waveform templates. However, we know that binaries with substantial eccentricities exist,
e.g. the Hulse-Taylor binary with an eccentricity of $e \sim 0.6$~\cite{weisberg-2016}.
Nonetheless, at the time this binary enters the detection band of ground-based GW detectors, it
will have circularized to a negligible $e \sim 10^{-5}$ and not be distinguishable from a circular
binary with current detector sensitivity~\cite{peters-1963,sun-2015,huerta-2018}. In particular, in
globular clusters and galactic nuclei, there are expected to be binaries with non-negligible
eccentricity ($e>0.1$) emitting detectable
GWs~\cite{samsing-2018,samsing-2014,oleary-2007,naoz-2013,park-2017,oleary-2009}. Hence, the
detection of GWs from eccentric compact binaries could provide important information on compact
object populations in globular clusters and galactic nuclei~\cite{antonini-2016}.

As soon as LISA is operational, it will be able to observe compact binaries in our galaxy emitting
GWs of much lower frequency. At this point they still are expected to have moderate
eccentricities~\cite{nelemans-2003,sesana-2017}. On the other hand, LISA should be able to detect
supermassive black hole binaries forming in the aftermath of galaxy mergers. Notably triple-induced
coalescences are expected to have large eccentricities that remain significant until
merger~\cite{blaes-2002,hoffman-2007,amaro-2010,bonetti-2018-1,bonetti-2018-2}.

The above anticipated prospects of future GW observations have motivated the development of 
eccentric waveform models. In the inspiraling phase one usually uses the post-Newtonian (PN)
formalism to model the dynamics of the binary. This introduces three distinct timescales. The first
two, the orbital and the periastron precession time scales, are associated with the conservative
dynamics and commonly described by the quasi-Keplerian
parametrization~\cite{damour-1985,memmesheimer-2004}. The third time scale appears when the
dissipative radiation-reaction effects are taken into
account~\cite{damour-2004,koenigsdoerffer-2006}. Several waveform models have been built using this
description of a binary~\cite{yunes-2009, cornish-2010, key-2011,huerta-2014,huerta-2017,
huerta-2018, gopakumar-2011, tanay-2016, hinder-2018, cao-2017, klein-2018}. In general, the
far-zone gravitational radiation field receives instantaneous and hereditary contributions. The
instantaneous part is determined by the state of its source at a given retarded time while the
hereditary part depends on the entire dynamical history of the source. In particular, the latter
contains tail and memory pieces.

In this work, we concentrate on the memory contributions to the waveform from eccentric binaries.
Normally we think of gravitational waves as oscillatory perturbations propagating on the background
metric at the speed of light. However, all GW sources are subject to the so-called
gravitational-wave memory effect, which manifests in a difference of the observed GW amplitudes at
late and early times:
\begin{align}
	\Delta h_\mem &= \lim_{t \rightarrow +\infty} h(t) - \lim_{t \rightarrow - \infty} h(t) \,.
\end{align}
In an ideal, freely falling GW detector the GW memory causes a permanent displacement after the GW
has passed. There are two main types of GW memory: The linear memory~\cite{zeldovich-1974}
originates from a net change in the time derivatives of the source multipole moments between early
and late times, present mainly in unbound (e.g. hyperbolic binary) systems. For bound systems the
linear memory is negligible, as long as the components were formed, captured or underwent mass loss
long before the GW-driven regime. The nonlinear memory, also called ``Christodoulou
memory"~\cite{christodoulou-1991,wiseman-1991,thorne-1992,blanchet-1992,payne-1983}, is a phenomenon
directly related to the nonlinearity of general relativity. It arises from GWs sourced by
previously emitted GWs. Since the nonlinear memory is not produced directly by the source but
rather by its radiation, it is present in all sources of GWs. From a more theoretical perspective,
the memory effect and its variants can be interpreted in terms of conserved charges at null infinity
and ``soft theorems"~\cite{strominger-2016,pasterski-2016}. Several methods to look for the memory
effect have been devised. PTAs would observe a sudden change in the pulse frequency of a
pulsar~\cite{seto-2009,vanhaasteren-2010,cordes-2012,wang-2015} and ground-based detectors like
LIGO---although not sensitive enough to the memory of a single event---could allow for a detection
from the accumulation of several events.~\cite{favata-2009-2,lasky-2016,mcneill-2017}.

For circular binaries the nonlinear, nonoscillatory memory contributions to the waveform were
computed at the 3PN order in Ref.~\cite{favata-2009}. Regarding eccentric binaries, the
leading-order zero-frequency or the so-called direct current (DC) memory terms were obtained in 
Ref.~\cite{favata-2011}. In this paper, we extend these computations to the 3PN level by computing
all terms coming from the memory contribution to the radiative mass multipoles. Note that this
yields not only the ``genuine'' DC memory, but also oscillatory contributions. In the circular
limit, the latter have been computed in Ref.~\cite{blanchet-2008}. Due to complicated hereditary
integrals, we calculate the memory contributions within a small-eccentricity expansion. We present
all of our results in modified harmonic (MH) gauge in terms of the post-Newtonian parameter $x = (
G m \bar{\omega} / c^3 )^{2/3}$ and the eccentricity $e = \eb_t$, with $\bar{\omega} =
(1+\bar{k})\bar{n}$ being the orbital frequency and $\bar{n} = 2 \pi / P$ the mean motion. With the
instantaneous contributions already available~\cite{mishra-2015}, and the tail and post-adiabatic
contributions computed in a companion paper~\cite{boetzel-2019}---hereafter called Paper I---this
work aims to complete the knowledge of the 3PN waveform valid during the early inspiral of
eccentric binary systems.

This paper is structured as follows. In~\cref{sec:prerequisites} we discuss how the nonlinear
memory arises from the gravitational-wave energy flux and how it can be computed by integrating this
flux over the binary's past history. In~\cref{sec:memcomp} we explicitly evaluate the past-history
integrals, which lead to two types of memory terms---DC memory and oscillatory memory---that are
discussed separately in~\cref{sec:DCmemory,sec:oscmemory}. We next combine our results with the
already available instantaneous and tail contributions and discuss the full 3PN waveform
in~\cref{sec:fullwaveform}. In~\cref{sec:summary} we give a brief summary and conclude our work.
Most expressions in this paper are presented only to leading order in eccentricity for convenience,
though we provide the complete results to $\bigO(e^6)$ in a supplemental \textit{Mathematica}
notebook~\cite{supplement}.

\section{Prerequisites} \label{sec:prerequisites}

\subsection{Memory contribution to the mass multipole moments}

Here we briefly state the essentials of the memory calculation. The conventions and notations used
are the same as those outlined in Sec.~II of Paper I.
 
The gravitational waveform polarizations can be uniquely decomposed into the spherical harmonic
modes $h^{\ell m}$ via
\begin{align}\label{eq:hlmmodes}
	h_+ - \ui h_\times &= \sum_{\ell=2}^{\infty} \sum_{m=-\ell}^{\ell} h^{\ell m} \, Y^{\ell
		m}_{-2} (\Theta, \Phi) \,,
\end{align}
where the basis is formed by the spin-weighted spherical harmonics $Y^{\ell m}_{-2} (\Theta, \Phi)$
and the amplitude modes
\begin{align}\label{eq:hlmofu}
	h^{\ell m} &= -\frac{G}{\sqrt{2} R c^{\ell+2}} \left( U^{\ell m} - \frac{\ui}{c} V^{\ell m} 
		\right)
\end{align}
are given in terms of radiative mass and current multipoles, $U^{\ell m}$ and $V^{\ell m}$. These
contain both instantaneous and hereditary parts. In the latter, we can further distinguish between
tail and memory contributions (some of which may actually be tail induced) at the 3PN order, by
schematically writing
\begin{subequations}
	\begin{align}
		U^{\ell m} &= U^{\ell m}_\inst + U^{\ell m}_\tail + U^{\ell m}_\mem + \delta U^{\ell m}\,,\\
		V^{\ell m} &= V^{\ell m}_\inst + V^{\ell m}_\tail + \delta V^{\ell m} \,,
	\end{align}
\end{subequations}
where $\delta U^{\ell m}$ and $\delta V^{\ell m}$ represent possible higher-order hereditary terms.
Note that there is no memory contribution to the radiative current-type
moments~\cite{blanchet-1992}. Now, employing the multipolar post-Minkowskian post-Newtonian (PN)
formalism, the radiative moments can be written in terms of the source moments. These
relations can be found in Sec.~III A of Ref.~\cite{mishra-2015} for the instantaneous parts, which
only require the knowledge of the source motion at a given moment in retarded time $T_R$, and in
Sec.~II B of Paper I for the hereditary parts, which involve integrals over the entire dynamical
past of the source.

The nonlinear memory may be expressed in terms of the time derivative of the gravitational waveform
by solving one component of Einstein's equations near future null infinity in Bondi
coordinates~\cite{strominger-2016, nichols-2017}. In this approach, the complex wave amplitude
$h_+-\ui h_\times$ is decomposed into even-parity and odd-parity pieces, the former being
parametrized by a scalar function of the retarded time $T_R$ and the angles $(\Theta, \Phi)$,
namely $\Phi_{\text{e}}(T_R,\Theta,\Phi)=\sum_{\ell\ge 0, |m|\le \ell} \Phi^{\ell m}_{\text{e}}(T_R)
Y^{\ell m}(\Theta,\Phi)$, where the $\Phi^{\ell m}_{\text{e}}(T_R)$ turn out to be equal to
$\sqrt{2(\ell-2)!/(\ell+2)!}\, U^{\ell m}$ with our conventions. The memory then manifests itself
as a low-frequency shift of those modes. Since this effect is sourced by GWs, the consequent change
for $U^{\ell m}_\mem$ is a functional of the gravitational-wave
``flux''\footnote{In~\cref{eq:energy flux}, at future null infinity the product of $R^2$ with the
term between brackets reduces to $N_{AB}N^{AB}/2$ in the notation of Ref.~\cite{nichols-2017}.}
\begin{align} \label{eq:energy flux}
	\frac{\ud E^\GW}{\ud t \, \ud \Omega} &\equiv \frac{c^3 R^2}{16 \pi G} \left(
		\dot{h}_+^2 + \dot{h}_\times^2 \right) \,.
\end{align}
More precisely, the nonlinear memory contribution to the radiative mass moment $U^{\ell m}(T_R)$ is
given by~\cite{strominger-2016}
\begin{align}\label{eq:Ulmem}
	U^{\ell m}_\mem &= \frac{32\pi}{c^{2-\ell}} \sqrt{\frac{(\ell -2)!}{2(\ell +2)!}}
		\int_{-\infty}^{T_R} \ud t \int \ud \Omega \frac{\ud E^\GW}{\ud t \ud \Omega} \bar Y^{\ell
		m}(\Omega) \,.
\end{align}
This formula was first shown to hold at quadratic order in $G$~\cite{blanchet-1992} before its
validity was extended to the general case (see also Ref.~\cite{blanchet-1987}, which indicates how
to perturbatively construct a radiative-type gauge in which the derivation proposed in
Ref.~\cite{blanchet-1992} can be adapted in principle to arbitrarily high orders). We will start
from~\cref{eq:Ulmem} to compute the memory contributions to the GW amplitude to the 3PN order.

Inserting the mode decomposition defined in~\cref{eq:hlmmodes} into~\cref{eq:energy flux}, we find
the GW energy flux in terms of the time derivatives of the $h^{\ell m}$ modes:
\begin{align}\label{eq:gwflux}
	\frac{\ud E^\GW}{\ud t \, \ud \Omega} = &\; \frac{c^3 R^2}{16 \pi G} \sum_{\ell'=2}^{\infty}
		\sum_{\ell''=2}^{\infty} \sum_{m'=-\ell'}^{\ell'} \sum_{m''=-\ell''}^{\ell''} 
		\dot{h}^{\ell' m'} \dot{\bar{h}}^{\ell'' m''} \no
		&\times Y^{\ell' m'}_{-2}(\theta,\phi) \bar Y^{\ell'' m''}_{-2}(\theta,\phi) \,.
\end{align}
We insert this expression in turn into~\cref{eq:Ulmem}. The time derivative of the memory
contribution to the mass multipole moment, $U^{\ell m(1)}_\mem = \ud U^{\ell m}_\mem/ \ud T_R$,
which is nothing but the memory contribution before integration over past history, may thus be
expressed as
\begin{align}\label{eq:Ulmmem}
	U^{\ell m(1)}_\mem =&\; \frac{c^{\ell+1} R^2}{G} \sqrt{\frac{2(\ell-2)!}{(\ell+2)!}}
		\sum_{\ell'=2}^{\infty} \sum_{\ell''=2}^{\infty} \sum_{m'=-\ell'}^{\ell'}
		\sum_{m''=-\ell''}^{\ell''} \no
		&\times G^{\ell \ell' \ell''}_{m m'm''} \dot{h}^{\ell'm'} \dot{\bar h}^{\ell''m''}\,,
\end{align}
where $G^{\ell \ell' \ell''}_{m m'm''}$ is the angular integral of a product of three spin-weighted
spherical harmonics,
\begin{align}\label{eq:angint}
	G^{\ell \ell' \ell''}_{m m' m''} = \int \ud \Omega \, \bar Y{^{\ell m}} \, Y^{\ell' m'}_{-2} \,
		\bar Y^{\ell'' m''}_{-2} \,.
\end{align}
Reference~\cite{NIST:DLMF} provides an explicit formula for this integral:
\begin{align}\label{eq:G}
	G^{\ell \ell' \ell''}_{m m'm''} =&\; (-1)^{m + m'} \sqrt{\frac{(2\ell +1)(2\ell' +1)(2\ell''
		+1)}{4\pi}} \no
		&\times
		\begin{pmatrix}
			\ell & \ell' & \ell'' \\
			0 & -2 & 2
		\end{pmatrix}
		\begin{pmatrix}
			\ell & \ell' & \ell'' \\
			-m & m' & -m''
		\end{pmatrix} \,.
\end{align}
The brackets denote the Wigner 3-$j$ symbols.

\subsection{Instantaneous and tail parts of the spherical harmonic modes} \label{sec:hlminsttail}

Remembering that the dominant modes correspond to the quadrupolar case $\ell = 2$, with $h^{2m} =
\bigO(c^{-4})$, we see from~\cref{eq:Ulmmem} that the memory integrands are of 2.5PN order.
However, as discussed below, in addition to oscillatory complex exponentials the $U^{\ell 0 
(1)}_\mem$ also contain nonoscillatory terms. Due to the integration over the past history, their
contributions at times $t\le T_R$ accumulate and enhance the result by a net factor $c^5$.
It follows that the leading memory effect in the polarizations actually arises at the relative
Newtonian order. Thus,~\cref{eq:Ulmmem} implies that, as an input for the computation of the
3PN-accurate $U^{\ell m(1)}_\mem$, we need \emph{a priori} all nonmemory $h^{\ell m}$ modes to 3PN
order. It is in fact not surprising that part of the waveform is required to calculate the full
waveform since the nonlinear memory originates from gravitational waves sourced by the energy flux
of gravitational waves emitted in the past, as shown by~\cref{eq:Ulmem}. Note that the contribution
from the memory to the memory itself turns out not to enter the waveform up to the 3PN order.
In Ref.~\cite{favata-2009} it was argued that for circular binaries these contributions would
appear at the 5PN level, though for eccentric binaries we find, by explicit calculation, that these
appear already at the 4PN order. This is due to additional oscillatory memory contributions that
will be discussed in~\cref{sec:oscmemory} below. However, for the present work, these
memory-of-memory terms can be safely ignored.

The instantaneous parts of the 3PN-accurate $h^{\ell m}$ modes describing inspiraling eccentric
binaries have been computed in Ref.~\cite{mishra-2015}. The tail contributions were derived in
Paper I, as well as the post-adiabatic corrections to the instantaneous contributions. The
instantaneous mode amplitudes from Ref.~\cite{mishra-2015} are written in terms of the
post-Newtonian parameter $x$, the time eccentricity $e_t$ and parametrized by the eccentric anomaly
$u$. They are valid for arbitrary eccentricities, while the tail contributions in Paper I are given
in a small-eccentricity expansion, parametrized by the mean anomaly $l$. The same will hold for the
memory parts. Inverting the 3PN-accurate Kepler equation by means of the solution developed in
Ref.~\cite{boetzel-2017}, the instantaneous terms can be parametrized by the mean anomaly as well.
Instead of restating the quasi-Keplerian parametrization and the phasing formalism describing the
dynamics of the binary, we refer the reader to Secs.~II C and II D of Paper I where those aspects
are summarized with the same conventions and notations.

The $h^{\ell m}$ modes including instantaneous, tail, and post-adiabatic contributions are given
in the following form:
\begin{align}\label{eq:hlm}
	h^{\ell m} &= \frac{8 Gm \nu}{c^2 R} x \sqrt{\frac{\pi}{5}} \ue^{-\ui m \psi} H^{\ell m} \,,
\end{align}
where the $H^{\ell m}$ are written in terms of the adiabatic post-Newtonian parameter $x \equiv
\xb$ and time eccentricity $e \equiv \eb_t$, and are parametrized by the angles $\xi$ and $\psi$.
See, for instance, Eq.~(76) of Paper I for the dominant mode ($h^{22}$) expression. The phase angles
$\xi$ and $\psi$ arise naturally when applying a certain shift to the time coordinate aimed at
eliminating the arbitrary constant $x_0$ appearing in both the instantaneous and tail
parts~\cite{blanchet-1996,arun-2004} through the redefinitions
\begin{subequations}
\begin{align}
	\xi &= l - \frac{3 G M}{c^3} n \ln \left( \frac{x}{x_0'} \right) \label{eq:xi} \,,\\
	\lambda_\xi &= \lambda - \frac{3 G M}{c^3} (1 + k) n \ln \left( \frac{x}{x_0'} \right) \,,
\end{align}
\end{subequations}
where $M = m (1 - \nu x / 2)$ denotes the Arnowitt-Deser-Misner (ADM) mass, $m = m_1 + m_2$ is the
total mass, $\nu = m_1 m_2 / m^2$ is the symmetric mass ratio, and $x_0'$ is related to $x_0$ by
\begin{align}
	\ln x_0' &= \frac{11}{18} - \frac{2}{3} \eg - \frac{4}{3} \ln 2 + \frac{2}{3} \ln x_0 \,,
\end{align}
with $\eg$ being Euler's constant. We refer to Appendix B of Paper I for the relations between the
orbital elements ($l$, $\lambda$, $\phi$) and their redefined counterparts ($\xi$, $\lambda_\xi$,
$\psi$).

\section{Computation of the nonlinear memory} \label{sec:memcomp}

\subsection{Memory contributions to the time derivative of the radiative moments}
\label{sec:Umemdots}

The computation of the memory contributions to the radiative mass multipole using~\cref{eq:Ulmmem}
involves products of the time derivatives of the $h^{\ell m}$ modes given in~\cref{eq:hlm}.
These are obtained by expressing $\psi$ in terms of $\xi$ and $\lambda_\xi$ and applying the
following time derivative operator:
\begin{align}
	\frac{\ud}{\ud t} &= n \left[\frac{\ud}{\ud \xi} + (1+k) \frac{\ud}{\ud \lambda_\xi} \right] +
		\frac{\ud x}{\ud t} \frac{\ud}{\ud x} + \frac{\ud e}{\ud t} \frac{\ud}{\ud e} \,,
\end{align}
where we have used the facts that $\ud \xi/\ud t = \ud l/\ud t = n$ and $\ud \lambda_\xi/\ud t =
\ud \lambda/\ud t = (1+k)n$ to the required PN order. The secular time evolution of $x$ and $e$ is
given, at leading order, by the formulas of Peters and Mathews~\cite{peters-1963,peters-1964}
\begin{subequations}\label[eqs]{eq:peters-mathews}
\begin{align}
	\frac{\ud x}{\ud t} =& \frac{c^3 \nu}{Gm} \frac{x^5}{(1-e^2)^{7/2}} \left(\frac{64}{5}
		+ \frac{584}{15} e^2 + \frac{74}{15} e^4 \right) \,,\\
	\frac{\ud e}{\ud t} =& -\frac{c^3 \nu}{Gm} \frac{e \, x^4}{(1-e^2)^{5/2}} \left(
		\frac{304}{15} + \frac{121}{15} e^2 \right) \,.
\end{align}
\end{subequations}
Note that they cause a 2.5PN correction, thus the leading order is sufficient here. When computing
the time derivatives of the amplitude modes $h^{\ell m}\sim x^{\ell/2}/c^2$, we have the following
leading-order PN scaling:
\begin{align}
	\dot{h}^{\ell m} \sim (\omega/c^2)\, x^{\ell/2} \sim c\, x^{\ell/2+ 3/2}\,.
\end{align}
As the dominant mode $\ell=2$ is of order $c\,x^{5/2}$, the knowledge of the waveform to 3PN order
requires modes up to $\ell=8$. According to this argument, the sums in~\cref{eq:Ulmmem} consisting
of products $\dot{h}^{\ell'm'} \dot{\bar h}^{\ell''m''}$ may be truncated at $\ell'=\ell''=8$.
Moreover, the appearance of the 3-$j$ symbols in~\cref{eq:G} imply some selection rules: the three
lower entries have to add up to zero, i.e., $m= m'-m''$. Since the mode products appearing
in~\cref{eq:Ulmmem} scale like
\begin{align}
	\dot{h}^{\ell'm'} \dot{\bar h}^{\ell''m''} \sim x^{n/2} \, \ue^{-\ui (m'-m'')\lambda_\xi} \,,
\end{align}
for some integer $n$, only memory modes with $m=0$ will contain DC terms, as was previously found
for circular orbits~\cite{favata-2009}. The scaling being the same as in that case, we have to
compute the $U^{\ell 0 (1)}_\mem$ up to $\ell=10$. On the other hand, a mode separation property
holds for planar orbits~\cite{kidder-2008,faye-2012}:the $h^{\ell m}$ only depend on the mass
(current) radiative moments if $\ell+m$ is even (odd). Thus, as there is no memory effect in the
current radiative moment, there is no memory effect when $\ell+m$ is odd.

As an example, we show the leading-order part of the $20$-mode up to $\bigO(e^2)$, which will
represent the dominant memory contribution:
\begin{align}
	U^{20 (1)}_\mem =&\; -\sqrt{\frac{\pi}{15}} \frac{c^5\nu^2}{G} x^5 \bigg(\frac{256}{7}
		+\frac{5008 e^2}{21}+\frac{768}{7} e\, \ue^{-\ui \xi} \no
		&+ \frac{768}{7} e\, \ue^{\ui \xi}+\frac{5176}{21} e^2\, \ue^{-2 \ui \xi}+\frac{5176}{21}
		e^2\, \ue^{2 \ui \xi}\bigg)\,.
\end{align}
We observe two different type of terms: oscillatory terms proportional to $\ue^{-\ui m'\xi}$, and
nonoscillatory ones, which give rise to the well-known leading-order DC memory.

As argued above, the $m \neq 0$ modes only contain oscillatory terms since they are proportional to
$\ue^{-\ui m \lambda_\xi}$. For instance, the leading order of the $22$-mode explicitly is
\begin{align}
	U^{22 (1)}_\mem = &-\sqrt{\frac{2\pi}{5}} \frac{c^5\nu^2}{G} x^5 \, \ue^{-2 \ui \lambda_\xi}
		\bigg( \frac{40}{7}e^2 -\frac{32}{21} e\, \ue^{-\ui \xi}\no
		&+\frac{32}{21} e\, \ue^{\ui \xi} -\frac{172}{21} e^2 \ue^{-2 \ui \xi} +\frac{52}{21} e^2
		\ue^{2 \ui \xi}\bigg)\,.
\end{align}
However, in~\cref{eq:Ulmmem} (used for the calculation of the $U^{\ell m(1)}_\mem$) there is no need
to average the mode products $\dot{h}^{\ell' m'}\dot{\bar{h}}^{\ell'' m''}$ over several
wavelengths. Other derivations of the memory effect, making use of the Isaacson gravitational-wave
stress-energy tensor~\cite{isaacson-1968}, resort to such a procedure. In Ref.~\cite{favata-2011},
which follows this approach, the orbital average entering the calculation of the leading-order
eccentric memory effectively removes the terms proportional to $\ue^{\ui m'\xi}$ in the
$U^{\ell 0(1)}_\mem$, so that only the terms yielding the DC memory are left over, while the
discarded pieces do not affect the amplitude of the DC memory. In the absence of the orbital
average, these pieces lead to small-amplitude oscillatory contributions to the waveform, which here
we will call oscillatory memory contributions. It would actually be difficult to introduce an
orbital average in the $m \neq 0$ modes because these terms oscillate not only on the orbital time
scale but also on the much longer precession time scale.

\subsection{DC memory}\label{sec:DCmemory}

The next step consists in evaluating the hereditary time integral
\begin{align}\label{eq:heredint}
	U^{\ell m}_\mem = \int_{-\infty}^{T_R} \ud t \, U^{\ell m(1)}_\mem\,.
\end{align}
To do so, we need a model for the secular evolution of the binary undergoing gravitational
radiation-reaction forces. The secular 3PN-order evolution equations of the orbital elements for a
quasielliptical, inspiraling binary were obtained in
Refs.~\cite{arun-2008-1,arun-2008-2,arun-2009}. This model is an idealization since it assumes that
the two components start at infinite separation and the orbital energy decreases solely due to the
emission of gravitational waves.

The explicit integrals appearing in~\cref{eq:heredint} are of two different types. The first one
consists of a product of $x$ and $e$, each with some power $p$ and $q$, respectively:
\begin{align}\label{eq:DCmemint}
	U^{\ell 0}_\memdc &\sim \int_{-\infty}^{T_R} \ud t\, x^p(t) \, e^q(t)\,.
\end{align}
The leading Newtonian order corresponds to $p=5$. The possible values of the integer $q$ range from
$0$---the quasi-circular limit---to the order of eccentricity expansion. These integrals give the
nonoscillatory contributions to the waveform, i.e., the DC memory. Note that, as argued above, these
terms are only present in the $m=0$ modes. The second type of integrals will lead to oscillatory
terms appearing at 1.5PN, 2.5PN, and 3PN order in the waveform. We will discuss these
in~\cref{sec:oscmemory}.

The strategy to evaluate the integral in~\cref{eq:DCmemint} is to express the PN parameter $x$ in
terms of the eccentricity $e$ and change the integration variable from time $t$ to $e$, so that
the integral runs from some initial eccentricity $e_i$ at early times to $e(T_R)$ at the current
retarded time:
\begin{align}\label{eq:DCint}
	U^{\ell 0}_\memdc \sim \int_{e_i}^{e(T_R)} \ud e\,\left(\frac{\ud e}{\ud t}\right)^{-1} x^p(e)
	\, e^q \,.
\end{align}
The time evolution of $x$ and $e$ due to radiation reaction is stated to leading order
in~\cref{eq:peters-mathews}. Here, we need the evolution equations up to 3PN order, which are
provided in~\cref{sec:evolveq}. We form the ratio of the two equations, thereby canceling the time
dependence, and expand the right-hand side in $x$ and $e$. This yields a differential equation with
the following structure:
\begin{align}\label{eq:dxde}
	\frac{\ud x}{\ud e} = &\; f_\mathrm{N}(e) \, x + f_1(e)\, x^2 + f_{1.5}(e) \,x^{5/2} \no
		&+f_2(e) \,x^3 + f_{2.5}(e) \,x^{7/2} + f_3(e, \ln x) \,x^4 \,.
\end{align}
Here, the $f_i(e)$ terms represent the coefficients of $x^{i+1}$ in the expansion of $\ud x/\ud e$,
with $f_{\mathrm{N}}=f_0$. To solve this differential equation, we search for the unknown function
$x(e)$ in the form of a perturbative expansion, according to
\begin{align}
	x =&\; x_\mathrm{N} + \epsilon \, x_1 + \epsilon^{3/2} \, x_{1.5} + \epsilon^2 \, x_2 
	+\epsilon^{5/2} \, x_{2.5} + \epsilon^3 \, x_3 \,,
\end{align}
\\
where $\epsilon$ is a formal parameter that allows one to keep track of the PN order. Inserting this
expansion into~\cref{eq:dxde} and identifying the coefficients of $\epsilon^i$ on the left- and the
right-hand sides of the resulting equation, we find the set of differential relations satisfied by
the post-Newtonian orders of $x$. This system can be straightforwardly solved by quadrature.
Putting the pieces together yields the PN parameter $x$ as a function of eccentricity. At leading
order in the PN and eccentricity expansion, we recover~\cite{damour-2004}
\begin{align}\label{eq:xofe}
	x(e) = x_0 \left(\frac{e_0}{e}\right)^{12/19}\,,
\end{align}
where $x_0$ is the value of $x$ at some reference eccentricity $e_0$. The full 3PN result to
leading order in eccentricity is provided in~\cref{sec:evolveq}. Note that for the expansion in
eccentricity to be valid, the eccentricity has to be small at all times, and hence $e_0$ has to be
small as well.

We are now in the position to insert the evolution equation for $e$ and the solution for $x(e)$
into~\cref{eq:DCint}. Expanding again in $x$ and $e$ yields elementary integrals,which must be
calculated. We then reexpress this result in terms of the time-dependent quantities $x$ and $e$ by
solving their relation~[\cref{eq:xofe,eq:xofe3}] for $x_0$ and reinsert the expression of this
quantity in terms of $x$ and $e$ into the calculated memory terms. A final Taylor expansion then
yields the DC memory pieces of the mass multipole moments.

We present the memory contributions to the spherical harmonic modes in the following form:
\begin{align}\label{eq:hlmmemory}
	h^{\ell m}_\mem =&\; - \frac{G}{\sqrt{2} c^{l+2} R} U^{\ell m}_\mem \no
		=&\; \frac{8 G m \nu}{c^2 R} x \sqrt{\frac{\pi}{5}} \ue^{-\ui m \psi} H^{\ell m}_\mem \,.
\end{align}
With this convention, the memory pieces directly add to the waveform modes stated in~\cref{eq:hlm}.
As the expressions are quite long, we present here only the $H^{20}_\memdc$ mode to 3PN and leading
order in eccentricity:
\begin{widetext}
\begin{subequations}\label{eq:h20mem}
\begin{align}
	H^{20}_\memdc =&\; -\frac{5}{14 \sqrt{6}} \left( H^{20}_\newt + x H^{20}_\pn{1} + x^{3/2}
		H^{20}_\pn{1.5} +x^{2} H^{20}_\pn{2} +x^{5/2} H^{20}_\pn{2.5} +x^{3} H^{20}_\pn{3}\right)
		\,,\\
	H^{20}_\newt =&\; 1 -\left( \frac{e}{e_i} \right)^{12/19} \,,\\
	H^{20}_\pn1 =&\;  -\frac{4075}{4032} +\frac{67 \nu}{48} +\left(\frac{e}{e_i} \right)^{12/19}
		\left( -\frac{2833}{3192} +\frac{197 \nu}{114} \right) +\left( \frac{e}{e_i}
		\right)^{24/19} \left( \frac{145417}{76608}-\frac{2849 \nu}{912}\right) \,,\\
	H^{20}_\pn{1.5} =&\; -\frac{377\pi}{228} \left(\frac{e}{e_i} \right)^{12/19}
		+\frac{377\pi}{228} \left( \frac{e}{e_i}\right)^{30/19} \,,\\
	H^{20}_\pn2 =&\; -\frac{151877213}{67060224} -\frac{123815 \nu}{44352} +\frac{205 \nu^2}{352}
		+\left(\frac{e}{e_i}\right)^{12/19} \left(\frac{358353209}{366799104} -\frac{738407
		\nu}{727776} -\frac{20597 \nu^2}{17328}\right)\no
		&+\left(\frac{e}{e_i}\right)^{24/19} \left(\frac{411966361}{122266368} -\frac{825950
		\nu}{68229} +\frac{561253 \nu^2}{51984}\right) \no
		&+\left(\frac{e}{e_i}\right)^{36/19} \biggl(-\frac{50392977379}{24208740864}+
		\frac{764295307 \nu}{48033216} -\frac{11654209\nu^2}{1143648}\biggr) \,,\\
	H^{20}_\pn{2.5} =&\; -\frac{253 \pi}{336} +\frac{253 \pi \nu}{84} +\left( \frac{e}{e_i}
		\right)^{12/19} \left( \frac{3763903 \pi}{7277760} +\frac{12788779 \pi \nu}{1819440}\right)
		+\left( \frac{e}{e_i} \right)^{24/19} \left(\frac{54822209 \pi}{8733312} -\frac{1074073 \pi
		\nu}{103968} \right) \no
		&+\left(\frac{e}{e_i}\right)^{30/19} \left(\frac{5340205 \pi}{1455552} -\frac{371345 \pi
		\nu}{51984} \right) +\left(\frac{e}{e_i}\right)^{42/19} \left(-\frac{424020733
		\pi}{43666560} +\frac{27049187 \pi \nu}{3638880} \right) \,,\\
	H^{20}_\pn{3} =&\;  -\frac{4397711103307}{532580106240} +\left(\frac{700464542023}{13948526592}
		-\frac{205 \pi^2}{96}\right) \nu +\frac{69527951 \nu^2}{166053888} +\frac{1321981
		\nu^3}{5930496} \no
		&+\left(\frac{e}{e_i}\right)^{12/19} \biggl[-\frac{4942027570449143}{96592876047360}
		-\frac{81025 \pi^2}{103968} +\frac{3317 \eg}{399} +\left( -\frac{10309531979}{7466981760}
		+\frac{3977 \pi^2}{3648}\right) \nu \no
		&+\frac{267351733 \nu^2}{82966464} +\frac{772583 \nu^3}{2222316} +\frac{12091 \ln 2}{5985}
		+\frac{78003 \ln 3}{5320} +\frac{3317 \ln x}{798}\biggr] +\frac{710645 \pi^2}{103968}
		\left( \frac{e}{e_i}\right)^{30/19} \no
		&+\left(\frac{e}{e_i}\right)^{24/19} \left(-\frac{31102835980319}{14049872879616}
		+\frac{279737759653 \nu}{167260391424} +\frac{26730466283 \nu^2}{1991195136}
		-\frac{397176241 \nu^3}{23704704}\right) \no
		&+\left(\frac{e}{e_i}\right)^{36/19} \left(-\frac{142763304914707}{25758100279296}
		+\frac{48901891428821 \nu}{919932152832} -\frac{400181473249 \nu^2}{3650524416}
		+\frac{2295879173 \nu^3}{43458624}\right) \no
		&+\left(\frac{e}{e_i}\right)^{48/19} \biggl[\frac{385621605844415513}{5740376633671680}
		-\frac{157405 \pi^2}{25992} -\frac{3317 \eg}{399} +\left(
		-\frac{49590995147570629}{478364719472640} +\frac{1271 \pi^2}{1216}\right) \nu \no
		&+\frac{3194536246463\nu^2}{34514049024} -\frac{1672948713 \nu^3}{45653504}-\frac{12091 \ln
		2}{5985} -\frac{78003 \ln 3}{5320} -\frac{3317 \ln x}{798} -\frac{6634}{2527} \ln \left(
		\frac{e}{e_i} \right) \biggr] \,.
\end{align}
\end{subequations}
All nonzero DC memory modes are presented to leading order in eccentricity in~\cref{sec:DCmemlist}
and to $\bigO(e^6)$ in the Supplemental Material~\cite{supplement}.

An important check is to take the circular limit of our calculated memory modes and compare to the
circular 3PN memory modes computed in Ref.~\cite{favata-2009}. To illustrate this fact, we take the
circular limit of the $20$-mode stated in Eqs.~(\ref{eq:h20mem}) by setting $e=0$ and find
\begin{align}
	H^{20}_\memdc = &-\frac{5}{14 \sqrt{6}} \Bigg\{1+ x\left(-\frac{4075}{4032}+ \frac{67
		\nu}{48}\right)+ x^2 \left(-\frac{151877213}{67060224}-\frac{123815 \nu}{44352}+\frac{205
		\nu^2}{352}\right) +x^{5/2} \left(-\frac{253 \pi}{336}+\frac{253 \pi \nu}{84}\right) \no
		&+x^3 \left[-\frac{4397711103307}{532580106240}
		+\left(\frac{700464542023}{13948526592}-\frac{205 \pi^2}{96}\right) \nu +\frac{69527951
		\nu^2}{166053888}+\frac{1321981 \nu^3}{5930496}\right] \Bigg\}\,,
\end{align}
\end{widetext}
in perfect agreement with Eq.~(4.3a) of Ref.~\cite{favata-2009}. The higher DC modes up to $\ell =
10$ in the circular limit are consistent with Eq.~(4.3) of Ref.~\cite{favata-2009} as well.
Moreover, we can check the leading eccentricity part at Newtonian order against Eq.~(2.35) in
Ref.~\cite{favata-2011}. They are found to be equal. Note that at Newtonian order the computation
of the DC memory is in principle possible for arbitrary eccentricities [see Eq.~(2.34) in
Ref.~\cite{favata-2011}]; however, this becomes difficult at higher PN orders, especially when tail
terms come into play.

\subsection{Oscillatory memory} \label{sec:oscmemory}

Before considering the oscillatory integrals, let us recall some properties of the nonlinear
memory. As mentioned at the beginning of~\cref{sec:hlminsttail}, the memory contribution to the
radiative mass multipole is formally of 2.5PN order. But due to the hereditary nature, the
nonoscillatory terms are raised by 2.5PN orders to appear already at the Newtonian level. From the
oscillatory terms we cannot expect the same behavior, due to the fact that the oscillations in the
remote past effectively cancel each other out. Thus, we expect that only the recent past will
contribute.

Examining the remaining oscillatory integrals, we notice that they are of the following form:
\begin{align}\label{eq:oscmemint}
	U^{\ell m}_\memosc &\sim \int_{-\infty}^{T_R} \ud t\, x^p(t) \, e^q(t) \, \ue^{\ui(s
	\lambda_\xi + r \xi)} \,.
\end{align}
Note that we have $s = -m$. Here we provide a formula to evaluate these integrals, its derivation
is presented in~\cref{sec:oscintegral}. Using the fact that $\lambda_\xi = (1+k)\xi$ and $\xi = nt$
to the required PN order as well as the notion that the integral is essentially given by the
contributions at the current time, we find
\begin{align}\label{eq:oscmem}
	U^{\ell m}_\memosc \sim -\frac{\ui}{n(r+s(1+k))} x^p\, e^q \, \ue^{\ui (s \lambda_\xi + r \xi)}
		\,,
\end{align}
where the time dependence on $T_R$ is not written explicitly. Expanding the denominator, we have to
distinguish between two different cases. The first applies if $r \neq -s$; we then find
\begin{align}\label{eq:fastosc}
	U^{\ell m}_\memosc \sim\; -\frac{\ui}{r+s} x^{p-3/2}\, e^q \, \ue^{\ui (s \lambda_\xi + r
		\xi)}\,.
\end{align}
Since $p=5$ at Newtonian order and the leading terms in the waveform are of order $x$, these
integrals lead to 2.5PN contributions to the waveform. As we have expected, these kinds of terms
oscillating on the orbital time scale keep their formal PN order, and we call them the fast
oscillatory memory.

On the other hand, for $r = -s$ we find
\begin{align}\label{eq:slowosc}
	U^{\ell m}_\memosc \sim& -\frac{\ui}{3s}\left[ x^{p-5/2} + x^{p-3/2}\left(-\frac{3}{2}+
		\frac{7\nu}{3}\right) \right]\no
		&\times e^q \, \ue^{\ui s (\lambda_\xi - \xi)} + \bigO(e^{q+2})\,.
\end{align}
This corresponds to terms that oscillate solely on the periastron precession time scale, and we
therefore call these terms the slow oscillatory memory. Because of the much slower oscillations,
they are enhanced by 1PN order (corresponding to the PN order of precession) and enter the waveform
at 1.5PN. Note also that in~\cref{eq:slowosc} eccentricity corrections of $\bigO(e^{q+2})$ appear,
whereas~\cref{eq:fastosc} would only be affected by eccentricity corrections starting at 3.5PN
order.

We provide the oscillatory memory contributions to the spherical harmonic modes in the same form as
for the DC memory, according to~\cref{eq:hlmmemory}. Besides the DC memory contribution, the
$20$-mode also contains fast oscillatory memory at 2.5PN:
\begin{align}
	H^{20}_\memosc =&\; \ui \frac{16\sqrt{6}}{7} \nu\,e\, x^{5/2} \biggl( -\ue^{-\ui \xi} +\ue^{\ui
		\xi} \no
		&-\frac{647}{576} e \,\ue^{-2 \ui \xi} +\frac{647}{576} e\, \ue^{2 \ui \xi} \biggr)\,.
\end{align}
Note that while the DC memory is purely real and therefore only affects the plus polarization (with
the usual conventions on the polarization triad), the oscillatory contributions influence
both polarizations.

In the $m \neq 0$ modes, only the oscillatory memory is present. For the dominant $22$-mode we find
\begin{subequations}
\begin{align}
	H&^{22}_\memosc = \ui  \, e^2 \nu \, \ue^{2\ui \xi} \biggl[ -\frac{13}{252} x^{3/2} \no
		&+ \left( \frac{697}{336} -\frac{865 \nu}{216} \right)x^{5/2} -\frac{29\pi}{126} x^{3}
		\biggr] \no
		&+ 21 \ui x^{5/2} e \nu  \left[ \frac{19}{6}e +\frac{4}{3} \ue^{-\ui \xi} - 4 \ue^{\ui \xi}
		+\frac{65}{24} e \ue^{-2\ui \xi}\right] \,.
\end{align}
\end{subequations}
Here the slow oscillatory part in the first and second lines is proportional to $\ue^{2 \ui \xi}$,
as we factored out $\ue^{-2\ui\psi}$ according to~\cref{eq:hlmmemory}. Three different PN orders of
slow oscillatory memory terms appear in this mode. The first one at 1.5PN arises from the
leading-order memory contribution to the radiative mass multipole at 2.5PN, so as expected it is
enhanced by one post-Newtonian order. At 2.5PN, there is the 1PN correction to the first term as
well as a part coming from the 1PN correction to the multipole. Finally, at 3PN there is a term
originating from the 1.5PN correction to the memory part of the multipole; this corresponds to
memory of the gravitational-wave tail. The terms in the third line correspond to fast oscillatory
memory entering at the 2.5PN level.

\section{Full 3PN eccentric waveform}\label{sec:fullwaveform}

In this section we summarize the results necessary to construct the full waveform for eccentric
binaries at third post-Newtonian order, including all instantaneous, hereditary and post-adiabatic
contributions, as described in Sec. V of Paper I. Rather than listing the lengthy expressions, we
give an overview at which PN order the individual terms enter the waveform and where they can be
found. Explicit expressions for all spherical harmonic modes are given in a supplemental
\textit{Mathematica} notebook~\cite{supplement}.

We present the waveform in terms of the secular evolving PN parameter $\xb$ and the time
eccentricity $\eb$, parametrized by the angles $\xi$ and $\psi$. We refer to Sec.~V C of Paper I
for their definition, and to Appendix B therein for various relations between the orbital elements
($l$, $\lambda$, $\phi$) and ($\xi$, $\lambda_\xi$, $\psi$). The secular evolution of the parameters
$\xb$ and $\eb$ is given in~\cref{sec:evolveq}. The spherical harmonic modes describing the
waveform are then written in the following form:
\begin{align}\label{eq:hlmfull}
	h^{\ell m} = \frac{8 G m \nu}{c^2 R} \xb \sqrt{\frac{\pi}{5}} \ue^{-\ui m \psi} H^{\ell m} \,.
\end{align}
Modes with $m<0$ can be calculated from
\begin{align}
	h^{\ell \, -m} = (-1)^\ell \, \bar h^{\ell m} \,.
\end{align}
In general, the individual modes can be split into three types of contributions:
\begin{align}
	H^{\ell m} = H^{\ell m}_\inst + H^{\ell m}_\hered + H^{\ell m}_\postad \,.
\end{align}
The instantaneous terms depend only on the instantaneous state of the source at a given retarded
time, with contributions at different orders relative to the leading order for each mode given as
\begin{align}
	H^{\ell m}_\inst =&\; (H^{\ell m}_\inst)_\lead + (H^{\ell m}_\inst)_\pn1 + (H^{\ell
		m}_\inst)_\pn{1.5} \nonumber\\
		&+ (H^{\ell m}_\inst)_\pn2 + (H^{\ell m}_\inst)_\pn{2.5} + (H^{\ell m}_\inst)_\pn3 \,.
\end{align}
These are given in terms of $x$, $e$, and $u$ in Eqs.~(5.09)--(5.11) and Eq.~(A1) of
Ref.~\cite{mishra-2015}. The parametrization in terms of $u$ has to be transformed to $\xi$ using
Eq.~(B2b) in Paper I.

The post-adiabatic contributions are introduced by radiation-reaction corrections to the
quasi-Keplerian parametrization, at relative 2.5PN order:
\begin{align}
	H^{\ell m}_\postad =&\; (H^{\ell m}_\postad)_\pn{2.5} \,.
\end{align}
They are given in Eqs.~(66)--(67) of Paper I.

The hereditary contributions, on the other hand, depend on the entire dynamical past of the binary
system. They can be further split into tail and memory parts:
\begin{align}
	H^{\ell m}_\hered = H^{\ell m}_\tail + H^{\ell m}_\mem \,.
\end{align}
For the tails we find contributions at different orders relative to the leading order for each mode
as
\begin{align}
	H^{\ell m}_\tail =&\; (H^{\ell m}_\tail)_\pn{1.5} + (H^{\ell m}_\tail)_\pn{2.5} + (H^{\ell
		m}_\tail)_\pn3 \,.
\end{align}
These are given in Eqs.~(47)--(48) of Paper I.

There is both DC memory and oscillatory memory:
\begin{align}
	H^{\ell m}_\mem = H^{\ell 0}_\memdc + H^{\ell m}_{\memosc} \,.
\end{align}
DC memory enters the waveform in the $m=0$ modes at all relative orders
\begin{align}
	H^{\ell 0}_\memdc =&\; (H^{\ell 0}_\memdc)_\lead + (H^{\ell 0}_\memdc)_\pn1 + (H^{\ell
		0}_\memdc)_\pn{1.5} \nonumber\\
		&+ (H^{\ell 0}_\memdc)_\pn2 + (H^{\ell 0}_\memdc)_\pn{2.5} + (H^{\ell 0}_\memdc)_\pn3 \,,
\end{align}
\\ 
while slow and fast oscillatory memory enter as
\begin{align}
	H^{\ell m}_\memosc =&\; (H^{\ell m}_{\textnormal{slow}\,\memosc})_\pn{1.5} + (H^{\ell
		m}_{\textnormal{slow} \,\memosc})_\pn{2.5} \nonumber\\
		&+ (H^{\ell m}_{\textnormal{slow} \,\memosc})_\pn3 + (H^{\ell m}_{\textnormal{fast}
		\,\memosc})_\pn{2.5} \,.
\end{align}
Slow oscillatory memory is due to the double-periodic nature of eccentric motion and is not present
in quasicircular binary systems. All memory modes are computed in this paper and are listed
in~\cref{sec:DCmemlist,sec:oscmemlist}.

As an example, we present here the dominant $H^{22}$ mode including all contributions to $\bigO(e)$:
\begin{widetext}
\begin{subequations}\label{eq:full_22-mode}
\begin{align}
	H^{22}_\newt =&\; 1 + \eb \bigg( \frac{1}{4} \ue^{-\ui\xi} + \frac{5}{4} \ue^{\ui\xi} \bigg)
		\,,\\
	H^{22}_\pn1 =&\; \xb \Bigg\{ -\frac{107}{42} + \frac{55 \nu}{42} + \eb \bigg[ \ue^{-\ui\xi}
		\left( -\frac{257}{168} + \frac{169 \nu}{168} \right) + \ue^{\ui\xi} \left( -\frac{31}{24}
		+ \frac{35 \nu}{24} \right) \bigg] \Bigg\} \,,\\
	H^{22}_\pn{1.5} =&\; \xb^{3/2} \Bigg\{ 2 \pi + \eb	\bigg[ \ue^{-\ui \xi} \left(
		\frac{11 \pi }{4} + \frac{27 \ui}{2} \ln \left( \frac{3}{2} \right) \right) + \ue^{\ui \xi}
		\left(\frac{13 \pi }{4} + \frac{3 \ui}{2} \ln(2) \right) \bigg] \Bigg\} \,,\\
	H^{22}_\pn2 =&\; \xb^2 \Bigg\{ -\frac{2173}{1512} - \frac{1069 \nu}{216} + \frac{2047
		\nu^2}{1512} + \eb \bigg[ \ue^{\ui\xi} \left( -\frac{2155}{252} - \frac{1655 \nu}{672} +
		\frac{371 \nu^2}{288} \right) \no
		&+ \ue^{-\ui\xi} \left( -\frac{4271}{756} - \frac{35131 \nu}{6048} + \frac{421 \nu^2}{864}
		\right) \bigg] \Bigg\} \,,\\
	H^{22}_\pn{2.5} =&\; \xb^{5/2} \Bigg\{ -\frac{107 \pi}{21} + \left( -24 \ui + \frac{34 \pi}{21}
		\right) \nu \no
		&+ \eb \bigg[ \ue^{\ui \xi} \bigg( -\frac{9 \ui}{2} + \frac{229 \pi}{168} + \left(
		-\frac{14579 \ui}{140} + \frac{61 \pi}{42} \right) \nu + \left( \frac{473 \ui}{28} -
		\frac{3 \ui \nu }{7} \right) \ln (2) \bigg) \no
		&+ \ue^{-\ui \xi} \bigg( -\frac{27 \ui}{2} -\frac{1081 \pi}{168} + \left( -\frac{1291
		\ui}{180} + \frac{137 \pi}{42} \right) \nu + \left( \frac{27 \ui}{4} + 9 \ui \nu \right)
		\ln \left( \frac{3}{2} \right) \bigg) \bigg] \Bigg\} \,,\\
	H^{22}_\pn3 =&\; \xb^3 \Bigg\{ \frac{27027409}{646800} + \frac{428 \ui \pi}{105} + \frac{2
		\pi^2}{3} - \frac{856 \eg}{105} + \left( -\frac{278185}{33264} + \frac{41 \pi^2}{96}
		\right) \nu - \frac{20261 \nu^2}{2772} + \frac{114635 \nu^3}{99792} \no
		&- \frac{1712 \ln(2)}{105} - \frac{428 \ln(\xb)}{105} \no
		&+ \eb \bigg[ \ue^{-\ui \xi} \bigg( \frac{219775769}{1663200} + \frac{749 \ui \pi}{60} +
		\frac{49 \pi^2}{24} - \frac{749 \eg}{30} + \left( -\frac{121717}{20790} - \frac{41
		\pi^2}{192}\right) \nu - \frac{86531 \nu^2}{8316} - \frac{33331 \nu^3}{399168} \no
		&+ \left( -\frac{2889}{70} + \frac{81 \ui \pi}{2}\right) \ln \left( \frac{3}{2} \right) -
		\frac{81}{2} \ln^2 \left( \frac{3}{2} \right) - \frac{749 \ln(2)}{15} - \frac{749
		\ln(\xb)}{60} \bigg) \no
		&+ \ue^{\ui \xi} \bigg( \frac{55608313}{1058400} + \frac{3103 \ui \pi}{420} + \frac{29
		\pi^2}{24} - \frac{3103 \eg}{210} + \left( -\frac{199855}{3024} + \frac{41 \pi^2}{48}
		\right) \nu -\frac{9967 \nu^2}{1008} + \frac{35579 \nu^3}{36288} \no
		&+ \left( -\frac{6527}{210} + \frac{3 \ui \pi}{2}\right) \ln(2) + \frac{3 \ln^2(2)}{2} -
		\frac{3103 \ln(\xb)}{420} \bigg) \bigg] \Bigg\} \,.
\end{align}
\end{subequations}
\end{widetext}
Note here the difference at 2.5PN order between~\cref{eq:full_22-mode} and Eq.~(76) of Paper I, due
to additional memory terms not yet considered in Paper I. Complete expressions for all modes to
$\bigO(e^6)$ are given in the Supplemental Material~\cite{supplement}.

By taking the quasicircular limit of our modes as described in Sec.~V E of Paper I, we can compare
the instantaneous, tail and (fast) oscillatory memory contributions of our waveform modes with
Ref.~\cite{blanchet-2008} and the DC memory terms with Ref.~\cite{favata-2009}. In all of them we
find perfect agreement.

\section{Brief summary}\label{sec:summary}

In this paper we computed the memory contribution to the gravitational waveform from nonspinning
compact binaries in eccentric orbits at the third post-Newtonian order. Our results complete the
previous work on the instantaneous parts~\cite{mishra-2015} and on the tail and post-adiabatic
contributions~\cite{boetzel-2019}. These waveforms form the basis for the construction of
increasingly accurate GW templates from binary systems in eccentric orbits.

There are two fundamentally different types of memory. DC memory is a slowly increasing,
nonoscillatory contribution to the gravitational-wave amplitude, entering at Newtonian order,
leading to a difference in the amplitude between early and late times. Oscillatory memory, on the
other hand, enters at higher PN orders as a normal periodic contribution. Due to the double-periodic
nature of the eccentric motion, slow oscillatory memory contributions on the periastron precession
time scale are enhanced by a factor of 1PN, and thus already enter the waveform at 1.5PN order. This
is unlike the quasicircular case, where oscillatory memory only enters at 2.5PN order.

\acknowledgments
We thank Marc Favata for an early review and useful comments. We also thank Maria Haney and 
Achamveedu Gopakumar for insightful discussions and comments, as well as Luc Blanchet for 
stimulating discussions. M.~E. and Y.~B. are supported by the Swiss National Science Foundation.
Y.~B. is supported by a Forschungskredit of the University of Zurich, grant no.~FK-18-084.

\appendix
\begin{widetext}
	
\section{Computation of the memory via the radiative mass multipoles} \label{sec:altmemory}

The computation of the nonlinear memory in the paper is done effectively via the GW energy flux
with the formula given in~\cref{eq:Ulmmem}. An alternative way is to directly compute the required
moments of the memory contribution to the radiative mass multipole. The leading-order memory piece
of the mass quadrupole moment contributes at 2.5PN, however, due to the hereditary integral the DC
terms are raised by 2.5PN orders such that they contribute at leading order in the waveform
polarization. Reference~\cite{faye-2015} lists the memory contributions up to 3.5PN. From this we
are able to compute the DC memory to 1PN accuracy. The hereditary integral enhances the slow
oscillatory memory terms by 1PN; therefore, by knowing the 3.5PN contribution to the mass moments
we find the leading-order 2.5PN terms contributing at 1.5PN and 2.5 PN in the waveform, and that
the 3PN terms appear at 2PN and 3PN and the 3.5PN terms at 2.5PN. However, what we miss are the 4PN
terms that appear in the waveform at 3PN level. On the other hand, the fast oscillatory memory is
not affected by the hereditary integral in its PN order, and we recover it at 2.5PN and 3PN. The
required memory contributions at 3.5PN to the radiative mass moments are
\begin{subequations}\label{eq:Umem}
\begin{align}
	U_{ij}^\mem(T_R) =&\; \frac{G}{c^5} \int_{-\infty}^{T_R} \ud \tau\, \left[ -\frac{2}{7}
		M^{(3)}_{a \langle i}\left(\tau \right) M^{(3)}_{j \rangle a}(\tau) \right] \no
		& +\frac{G}{c^7} \int_{ -\infty}^{T_R} \ud \tau \biggl[-\frac{5}{756} M^{(4)}_{a b}(\tau)
		M^{(4)}_{i j a b}(\tau) -\frac{32}{63} S^{(3)}_{a \langle i}(\tau) S^{(3)}_{j \rangle
		a}(\tau) \no
		& + \varepsilon_{a b \langle i} \left(\frac{5}{42} S^{(4)}_{j \rangle bc}(\tau)
		M^{(3)}_{ac}(\tau) -\frac{20}{189} M^{(4)}_{j \rangle bc}(\tau) S^{(3)}_{ac}(\tau) \right)
		\biggr]\,,\\
	U_{ijk}^\mathrm{mem}(T_R) =&\; \frac{G}{c^5} \int_{-\infty}^{T_R} \ud \tau \left[-\frac{1}{3}
		M^{(3)}_{a \langle i}(\tau) M^{(4)}_{jk \rangle a}(\tau) -\frac{4}{5} \varepsilon_{ab
		\langle i} M^{(3)}_{j a}(\tau) S^{(3)}_{k \rangle b}(\tau)\right]\,,\\
	U_{ijkl}^\mathrm{mem}(T_R) =&\; \frac{G}{c^3} \int_{-\infty}^{T_R} \ud \tau \left[\frac{2}{5}
		M^{(3)}_{\langle ij}(\tau) M^{(3)}_{kl \rangle}(\tau)\right] \no
		&+\frac{G}{c^5} \int_{-\infty}^{T_R} \ud \tau \biggl[\frac{12}{55} M^{(4)}_{a \langle
		i}(\tau) M^{(4)}_{jkl \rangle a}(\tau) -\frac{14}{99} M^{(4)}_{a \langle ij}(\tau)
		M^{(4)}_{k l\rangle a}(\tau) + \frac{32}{45} S^{(3)}_{\langle ij}(\tau) S^{(3)}_{k
		l\rangle}(\tau) \no
		& + \varepsilon_{a b \langle i} \left(-\frac{4}{5} M^{(3)}_{ja}(\tau) S^{(4)}_{kl \rangle
		b}(\tau) +\frac{32}{45} S^{(3)}_{ja}(\tau) M^{(4)}_{kl \rangle b}(\tau) \right) \biggr]\,,\\
	U_{ijklm}^\mathrm{mem}(T_R) =&\; \frac{G}{c^3} \int_{-\infty}^{T_R} \ud\tau \left[\frac{20}{21}
		M^{(3)}_{\langle ij}(\tau) M^{(4)}_{klm \rangle}(\tau)\right] \,,\\
	U_{ijklmn}^\mathrm{mem}(T_R) =&\; \frac{G}{c^3} \int_{-\infty}^{T_R} \ud \tau \left[\frac{5}{7}
		M^{(4)}_{\langle ijk}(\tau) M^{(4)}_{lmn \rangle}(\tau) -\frac{15}{14} M^{(3)}_{\langle
		ij}(\tau) M^{(4)}_{klmn \rangle}(\tau)\right]\,.
\end{align}
\end{subequations}
Note that the symmetric trace-free (STF) projection $\langle \dots \rangle$ only applies to the
free indices $ijk...$ The integrand in those equations consists of products of canonical mass and
current moments, $M^{(n)}_{L}(\tau)$ and $S^{(n)}_{L}(\tau)$,  and the superscript in brackets
stands for the $n{\mathrm{th}}$ derivative with respect to $\tau$. The canonical moments are
related by a gauge transformation to the source moments $I_L$ and $J_L$ along with some more gauge
moments that enter at 2.5PN in the $\delta I_L,\; \delta J_L$ terms,
\begin{subequations}
\begin{align}
	M_L &= I_L + G \delta I_L + \bigO(G^2) \,,\\
	S_L &= J_L + G \delta J_L + \bigO(G^2) \,.
\end{align}
\end{subequations}
For our purpose of calculating the memory contribution to next-to-leading order, we only need the
1PN part of the source moments. Here we list the relevant source moments at 1PN for two nonspinning
compact objects in general orbits~\cite{mishra-2015}. The source moments are written in terms of
$x_i$ and $v_i$, which denote the binary's relative separation and relative velocity. Moreover, $r$
is the distance between the two objects, and thus $r = |\boldsymbol{x}|$ and $\dot{r}$ is the
radial velocity. For the mass quadrupole moment we have
\begin{align}
	I_{ij} &= \nu m \left[ A_1 x_{\langle i} x_{j \rangle} + A_2 \frac{r \dot{r}}{c^2} x_{\langle
		i} x_{j \rangle} + A_3 \frac{r^2}{c^2} v_{\langle i} v_{j \rangle} \right]\,,
\end{align}
where
\begin{subequations}
\begin{align}
	A_1 &= 1 + \frac{1}{c^2} \left[v^2\left(\frac{29}{42} -\frac{29\nu}{14}\right) +\frac{G\,m}{r}
		\left( -\frac{5}{7} +\frac{8\nu}{7}\right)\right]\,,\\
	A_2 &= -\frac{4}{7}+ \frac{12 \nu}{7}\,,\\
	A_3 &= \frac{11}{21} - \frac{11 \nu}{7}\,.
\end{align}
\end{subequations}
The 1PN mass octupole is
\begin{align}
	I_{ijk} &= - \nu m \Delta \left[B_1 x_{\langle ijk \rangle} +B_2 \frac{r \dot{r}}{c^2}
		x_{\langle ij} v_{k \rangle} + B_3 \frac{r^2}{c^2} x_{\langle i} v_{jk \rangle} \right]\,,
\end{align}
where
\begin{subequations}
\begin{align}
	B_1 &= 1+ \frac{1}{c^2} \left[v^2\left(\frac{5}{6} -\frac{19\nu}{6}\right) +\frac{G\,m}{r}
		\left( -\frac{5}{6} +\frac{13\nu}{6}\right)\right]\,,\\
	B_2 &= - (1-2\nu)\,,\\
	B_3 &= 1 - 2 \nu\,,
\end{align}
\end{subequations}
and $\Delta = (m_1 - m_2) / m$ is the mass difference ratio. Moreover, we need also the
leading-Newtonian-order part of the mass hexadecapole,
\begin{align}
	I_{ijkl} = \nu m\, x_{\langle ijkl \rangle} \left(1 - 3 \nu \right)\,.
\end{align}
From the current source moments we need the quadrupole, which is
\begin{align}
	J_{ij} &= - \nu m \Delta \left[C_1 \varepsilon_{ab \langle i} x_{j \rangle a} v_{b} + C_2
		\frac{r \dot{r}}{c^2} \varepsilon_{ab \langle i} v_{j \rangle b} x_{a} \right]\,,
\end{align}
where,
\begin{subequations}
\begin{align}
	C_1 &= 1 + \frac{1}{c^2}\left[v^2 \left(\frac{13}{28}-\frac{17\nu}{7}\right)+ \frac{G \, m}{r}
		\left( \frac{27}{14} + \frac{15\nu}{7}\right)\right] \,,\\
	C_2 &= \frac{5}{28}(1-2\nu) \,,
\end{align}
\end{subequations}
and finally the leading order of the current octupole is
\begin{align}
	J_{ijk} &= \nu m \, \varepsilon_{ab \langle i} x_{jk \rangle a}v_{b} (1-3\nu) \,.
\end{align}
Having the source moments in hand (and thus in our case also the canonical moments), we can
calculate the products of time derivatives of the canonical moments occurring in the integrands
of~\cref{eq:Umem}. Before treating the hereditary integral, we transform from the STF moments
$U_L^\mem$ computed here to the scalar version of the radiative mass moments using Eq.~(4) of Paper
I. These are the same moments that we find when computing the memory with~\cref{eq:Ulmmem}. The
hereditary integral is evaluated in the same way as described in~\cref{sec:DCmemory,sec:oscmemory}.
Using this method, we find the 1PN DC memory and the 1PN oscillatory memory. Be aware that the DC
memory appears in the waveform at leading Newtonian order, while the first slow oscillatory memory
terms appear at 1.5PN and the fast oscillatory memory at 2.5PN.

This method of computing the memory contribution serves as a check. We can compare the relative 1PN
pieces of the DC and oscillatory memory calculated before and here, and they are found to be in
perfect agreement.

\section{Radiation-reaction evolution equations} \label{sec:evolveq}

In this appendix we provide the secular 3PN-accurate evolution equations for $x$ and
$e$~\cite{arun-2008-1,arun-2008-2,arun-2009} in (MH) gauge. The instantaneous terms are exact,
whereas the eccentricity enhancement functions appearing in the hereditary contributions are given
in an eccentricity expansion. We begin by listing the pieces needed for the evolution of $x$:
\begin{align}
	\frac{\ud x}{\ud t} = \frac{2 c^3 \nu x^5}{3 G m} \left(\mathcal{X}_\newt+ x \mathcal{X}_\pn1 +
		x^2 \mathcal{X}_\pn2 + x^3\mathcal{X}_\pn3 +\mathcal{X}_\hered\right)\,,
\end{align}
where
\begin{subequations}
\begin{align}
	\mathcal{X}_\newt =&\; \frac{1}{\left(1-e^2\right)^{7/2}} \left\{ \frac{96}{5}+\frac{292 
		e^2}{5} +\frac{37 e^4}{5}\right\}\,,\\
	\mathcal{X}_\pn1 =&\; \frac{1}{\left(1-e^2\right)^{9/2}}\left\{ -\frac{1486}{35} 
	-\frac{264 \nu}{5} +e^2 \left(\frac{2193}{7}-570 \nu \right) +e^4 
		\left(\frac{12217}{20}-\frac{5061 \nu}{10}\right) +e^6 \left(\frac{11717}{280}-\frac{148 
		\nu}{5}\right) \right\}\,,\\
	\mathcal{X}_\pn2= &\; \frac{1}{\left(1-e^2\right)^{11/2}}\biggl\{ -\frac{11257}{945} 
		+\frac{15677 \nu}{105} +\frac{944 \nu^2}{15} + e^2 \left(-\frac{2960801}{945} -\frac{2781 
		\nu}{5} + \frac{182387 \nu^2}{90}\right) \no
		&+e^4 \left(-\frac{68647}{1260}-\frac{1150631 \nu}{140}+\frac{396443 \nu^2}{72}\right) +e^6 
		\left(\frac{925073}{336}-\frac{199939 \nu}{48} +\frac{192943 \nu^2}{90}\right) \no
		&+e^8 \left(\frac{391457}{3360}-\frac{6037 \nu}{56}+\frac{2923 \nu^2}{45}\right) 
		+\sqrt{1-e^2} \biggl[ 48-\frac{96 \nu}{5} +e^2 \left(2134 -\frac{4268 \nu}{5}\right) \no
		&+e^4 \left(2193-\frac{4386 \nu}{5}\right) +e^6 \left(\frac{175}{2}-35 \nu \right)\biggr] 
		\biggr\}\,,\\
	\mathcal{X}_\pn3 =&\; \frac{1}{\left(1 -e^2\right)^{13/2}} \biggl\{ \frac{614389219}{148500} 
		+\left(-\frac{57265081}{11340}+\frac{369 \pi^2}{2}\right) \nu -\frac{16073 \nu^2}{140}- 
		\frac{1121 \nu^3}{27}\no
		&+e^2 \left(\frac{19769277811}{693000} +\left(\frac{66358561}{3240} +\frac{42571 
		\pi^2}{80}\right) \nu -\frac{3161701 \nu^2}{840} - \frac{1287385 \nu^3}{324}\right)\no
		&+e^4 \left(-\frac{3983966927}{8316000} + \left(\frac{6451690597}{90720} -\frac{12403 
		\pi^2}{64}\right) \nu + \frac{34877019 \nu^2}{1120}- \frac{33769597 \nu^3}{1296}\right)\no
		&+e^6 \left(-\frac{4548320963}{5544000} +\left(-\frac{59823689}{4032} -\frac{242563 
		\pi^2}{640}\right) \nu +\frac{411401857 \nu^2}{6720} -\frac{3200965 \nu^3}{108}\right)\no
		&+e^8 \left(\frac{19593451667}{2464000} + \left(-\frac{6614711}{480} -\frac{12177 
		\pi^2}{640}\right) \nu +\frac{92762 \nu^2}{7} -\frac{982645 \nu^3}{162}\right) \no
		&+e^{10} \left(\frac{33332681}{197120} - \frac{1874543 \nu}{10080} +\frac{109733 
		\nu^2}{840} -\frac{8288 \nu^3}{81} \right) \no
		&+\sqrt{1-e^2} \biggl[\left( -\frac{1425319}{1125} +\left(\frac{9874}{105} -\frac{41 
		\pi^2}{10}\right) \nu + \frac{632 \nu^2}{5}\right) \no
		&+e^2 \left(\frac{933454}{375} + \left(-\frac{2257181}{63} +\frac{45961 \pi^2}{240}\right) 
		\nu +\frac{125278 \nu^2}{15} \right) \no
		&+e^4 \left(\frac{840635951}{21000} + \left(-\frac{4927789}{60} +\frac{6191 
		\pi^2}{32}\right)\nu + \frac{317273 \nu^2}{15}\right)\no
		&+e^6 \left(\frac{702667207}{31500} + \left(-\frac{6830419}{252}+\frac{287 
		\pi^2}{960}\right) 
		\nu +\frac{232177 \nu^2}{30}\right) + e^8 \left(\frac{56403}{112} -\frac{427733 
		\nu}{840}+\frac{4739 \nu^2}{30} \right)\biggr] \no
		&+\log \biggl[\frac{x \left(1+ \sqrt{1-e^2}\right)}{x_0 \left(2 \left(1 
		-e^2\right)\right)}\biggr] \left(\frac{54784}{175} +\frac{465664 e^2}{105} +\frac{4426376 
		e^4}{525} +\frac{ 1498856 e^6}{525} + \frac{31779 e^8}{350}\right)\biggr\}\,,\\
	\mathcal{X}_\hered =&\; \frac{96}{5} \biggl\{4 \pi x^{3/2} \varphi (e) +\pi  x^{5/2} \left[ 
		-\frac{4159}{672} \psi_\omega (e) - \frac{189}{8} \nu \, \zeta_\omega (e) \right] \no
		&+x^3 \left[-\frac{116761}{3675} \kappa(e) + \left(\frac{16 \pi^2}{3} -\frac{1712 
		\eg}{105} - \frac{1712}{105} \log \left(\frac{4 x^{3/2}}{x_0}\right)\right)
		F(e)\right]\biggr\}\,.
\end{align}
\end{subequations}
The helper functions appearing in the hereditary contribution are given by
\begin{subequations}
\begin{align}
	\psi_\omega(e) =&\; \frac{1344}{4159} \frac{1}{\left(1 -e^2\right)^{3/2}} \left[\sqrt{1-e^2}
		\left(1-5 e^2\right) \varphi (e) -4 \tilde{\varphi}(e)\right] + \frac{8191}{4159}
		\psi(e)\,, \\
	\zeta_\omega(e) =&\; \frac{583}{567} \zeta(e) - \frac{16}{567} \varphi (e)\,.
\end{align}
\end{subequations}
The various enhancement functions appearing in these equations are listed below.

Next we state the evolution equation for the eccentricity. Note that we observed errors in the 2PN-
and 3PN-order expressions in Eqs.~(C10) and (C11) of Ref.~\cite{arun-2009}. These are likely due to
the fact that only the relation between $e^\mathrm{MH}$ and $e^\mathrm{ADM}$ was inserted, but one
also has to transform $\ud e^\mathrm{ADM}/\ud t$ to $\ud e^\mathrm{MH} / \ud t$,
\begin{align}
	\frac{\ud e}{\ud t} = - \frac{c^3 \nu e x^4}{G m} \left(\mathcal{E}_\mathrm{N} + x 
		\mathcal{E}_{\mathrm{1PN}} + x^2 \mathcal{E}_{\mathrm{2PN}} + x^3\mathcal{E}_{\mathrm{3PN}} 
		+ \mathcal{E}_\mathrm{hered} \right)\,,
\end{align}
where
\begin{subequations}
\begin{align}
	\mathcal{E}_\newt =&\; \frac{1}{\left(1-e^2\right)^{5/2}} \left\{ \frac{304}{15}+\frac{121 
		e^2}{15} \right\}\,,\\
	\mathcal{E}_\pn1 =&\; \frac{1}{\left(1-e^2\right)^{7/2}}\left\{ -\frac{939}{35} 
		-\frac{4084 \nu}{45} +e^2 \left(\frac{29917}{105} - \frac{7753}{30} \nu \right) +e^4 
		\left(\frac{13929}{280} - \frac{1664 \nu}{45}\right) \right\}\,,\\
	\mathcal{E}_\pn2 =&\; \frac{1}{\left(1 -e^2\right)^{9/2}} \biggl\{ -\frac{949877}{1890} 
		+\frac{18763 \nu}{42} +\frac{752 \nu^2}{5} + e^2 \left(-\frac{3082783}{2520} -\frac{988423 
		\nu}{840} +\frac{64433 \nu^2}{40} \right)\no
		&+e^4 \left(\frac{23289859}{15120} - \frac{13018711 \nu}{5040} +\frac{127411 \nu^2}{90}
		\right) +e^6\left(\frac{420727}{3360} -\frac{362071\nu}{2520} +\frac{821\nu^2}{9}\right)\no
		&+\sqrt{1-e^2} \biggl[ \frac{1336}{3} - \frac{2672 \nu}{15} +e^2 \left(\frac{2321}{2}
		-\frac{2321 \nu}{5}\right) +e^4 \left(\frac{565}{6} - \frac{113 \nu}{3}\right)\biggr]
		\biggr\}\,, \\
	\mathcal{E}_\pn3 =&\; \frac{1}{\left(1 -e^2\right)^{11/2}} \biggl\{ \frac{54208557619}{6237000} 
		+\left(\frac{50099023}{113400} + \frac{779 \pi^2}{10}\right) \nu -\frac{4088921
		\nu^2}{2520} - \frac{61001 \nu^3}{486} \no
		&+e^2 \left(\frac{46226320013}{6237000} + \left(\frac{28141879}{900} -\frac{139031  
		\pi^2}{960} \right) \nu -\frac{21283907 \nu^2}{3024} - \frac{86910509 \nu^3}{19440} 
		\right) \no
		&+e^4 \left(-\frac{116987170177}{16632000} + \left(\frac{11499615139}{907200} -\frac{271871 
		\pi^2}{1920}\right) \nu +\frac{61093675 \nu^2}{4032}- \frac{2223241 \nu^3}{180} \right) \no
		&+e^6 \left(\frac{5891934893}{1232000} + \left( -\frac{5028323}{560} -\frac{6519 
		\pi^2}{640}\right) \nu +\frac{24757667 \nu^2}{2520} - \frac{11792069 \nu^3}{2430}\right) \no
		&+e^8 \left( \frac{302322169}{1774080} -\frac{1921387 \nu}{10080} + \frac{41179 \nu^2}{216} 
		-\frac{193396 \nu^3}{1215} \right) \no
		&+\sqrt{1-e^2} \biggl[-\frac{22713049}{15750} +\left(-\frac{5526991}{945} +\frac{8323 
		\pi^2}{180}\right) \nu +\frac{54332 \nu^2}{45} \no
		&+e^2 \left( \frac{89395687}{7875} +\left( -\frac{38295557}{1260} +\frac{94177 
		\pi^2}{960}\right) \nu +\frac{681989 \nu^2}{90} \right) \no
		&+e^4 \left(\frac{5321445613}{378000} +\left( -\frac{26478311}{1512} +\frac{2501 
		\pi^2}{2880} \right) \nu +\frac{225106 \nu^2}{45}\right) \no
		&+e^6 \left(\frac{186961}{336} -\frac{289691 \nu}{504} +\frac{3197 \nu^2}{18} 
		\right)\biggr] +\frac{730168}{23625\left(1 +\sqrt{1 - e^2}\right)} \no
		&+\frac{304}{15} \left(\frac{82283}{1995} +\frac{297674}{1995} e^2 +\frac{1147147}{15960} 
		e^4 +\frac{61311}{21280} e^6\right) \ln \biggl[\frac{x \left(1 +\sqrt{1-e^2}\right)}{2 x_0 
		\left(1-e^2\right)} \biggr] \biggr\}\,,\\
	\mathcal{E}_\hered =&\; -\frac{32}{5} \biggl\{-\frac{985}{48} \pi  x^{3/2} \varphi_e(e) +\pi 
		x^{5/2} \biggl[ \frac{55691}{1344} \psi_e(e) + \frac{19067}{126} \nu \zeta_e(e) \biggr] \no
		&+x^3 \biggl[ \left(\frac{89789209}{352800} -\frac{87419 \ln 2}{630} +\frac{78003 \ln 
		3}{560}\right) \kappa_e(e) \no
		&-\frac{769}{96} \left(\frac{16 \pi^2}{3} -\frac{1712 \eg}{105} 
		-\frac{1712}{105} \ln \left(\frac{4 x^{3/2}}{x_0}\right)\right) F_e(e)\biggr] \biggr\}\,.
\end{align}
\end{subequations}
The additional functions in the hereditary contribution are
\begin{subequations}
\begin{align}
	\varphi_e(e) =&\; \frac{192}{985} \frac{\sqrt{1-e^2}}{e^2} \left[\sqrt{1-e^2} \varphi(e) 
		-\tilde{\varphi}(e)\right]\,,\\
	\psi_e(e) =&\; \frac{18816}{55691} \frac{1}{e^2 \sqrt{1-e^2}} \left[\sqrt{1-e^2} \left(1 
		-\frac{11 e^2}{7}\right) \varphi (e) -\left(1-\frac{3}{7} e^2\right) \tilde{\varphi}(e) 
		\right] \no
		&+\frac{16382}{55691} \frac{\sqrt{1-e^2}}{e^2} \left[\sqrt{1-e^2} \psi (e) 
		-\tilde{\psi}(e)\right]\,,\\
	\zeta_e (e) =&\; \frac{924}{19067} \frac{1}{e^2 \sqrt{1-e^2}} \left[-\left(1-e^2\right)^{3/2} 
		\varphi (e) + \left(1-\frac{5}{11} e^2\right) \tilde{\varphi}(e)\right] + 
		\frac{12243}{76268} \frac{\sqrt{1-e^2}}{e^2} \left[ \sqrt{1-e^2} \zeta (e) 
		-\tilde{\zeta}(e) \right]\,,\\
	\kappa_e (e) =&\; \frac{\sqrt{1-e^2}}{e^2} \left[ \sqrt{1-e^2} \kappa (e) 
		-\tilde{\kappa}(e)\right] \left(\frac{769}{96}-\frac{3059665}{700566} \ln 2 
		+\frac{8190315}{1868176}  \ln 3\right)^{-1}\,,\\
	F_e (e) =&\; \frac{96}{769} \frac{\sqrt{1-e^2}}{e^2} \left[\sqrt{1-e^2} F(e) - 
		\tilde{F}(e)\right]\,.
\end{align}
\end{subequations}
The eccentricity enhancement functions arise from hereditary contributions to the energy flux
(nontilde) and the angular momentum flux (tilde). Most of them do not admit closed forms and have
to be computed numerically or in a small-eccentricity expansion. Here we list them in an
eccentricity expansion to $\bigO(e^6)$:
\begin{subequations}
\begin{align}
	\varphi(e) =&\; 1+\frac{2335}{192} e^2 +\frac{42955}{768} e^4 +\frac{6204647}{36864} e^6\,,\\
	\tilde{\varphi}(e) =&\; 1+\frac{209}{32}  e^2+\frac{2415}{128} e^4 +\frac{730751}{18432} 
		e^6\,,\\
	\psi(e) =&\; 1-\frac{22988}{8191} e^2 -\frac{36508643}{524224} e^4 -\frac{1741390565}{4718016}  
		e^6 \,,\\
	\tilde{\psi}(e) =&\; 1-\frac{17416}{8191} e^2 -\frac{14199197}{524224} e^4 
		-\frac{467169215}{4718016} e^6 \,,\\
	\kappa(e) =&\; 1 +e^2 \left(\frac{62}{3} -\frac{4613840}{350283} \ln 2 
		+\frac{24570945}{1868176} \ln 3 \right) + e^4 \left(\frac{9177}{64} 
		+\frac{271636085}{1401132} \ln 2 -\frac{466847955}{7472704} \ln 3 \right) \no
		&+e^6\left( \frac{76615}{128} -\frac{4553279605}{2802264} \ln 2 
		+\frac{14144674005}{119563264} \ln 3 +\frac{914306640625}{1076069376} \ln 5 \right) \,,\\
	\tilde{\kappa}(e) =&\; 1 +e^2 \left(\frac{389}{32} - \frac{2056005}{233522} \ln 2 
		+\frac{8190315}{934088}\ln 3\right) +e^4 \left(\frac{3577}{64} + \frac{50149295}{467044} 
		\ln 2 -\frac{155615985}{3736352} \ln 3\right) \no
		&+e^6 \left( \frac{43049}{256} -\frac{12561332945}{16813584} \ln 2 
		+\frac{4709431125}{59781632} \ln 3 +\frac{182861328125}{538034688} \ln 5 \right) \,,\\
	\zeta(e) =&\; 1 +\frac{1011565}{48972} e^2 +\frac{106573021}{783552} e^4 
		+\frac{456977827}{854784} e^6 \,,\\
	\tilde{\zeta}(e) =&\; 1 +\frac{102371}{8162} e^2 +\frac{14250725}{261184} e^4 + 
		\frac{722230667}{4701312} e^6\,,\\
	F(e) =&\; 1+ \frac{62}{3} e^2 +\frac{9177}{64} e^4 +\frac{76615}{128} e^6 \,,\\
	\tilde{F}(e) =&\; 1 +\frac{389}{32} e^2 +\frac{3577}{64} e^4 +\frac{43049}{256} e^6 \,.
\end{align}
\end{subequations}
By dividing the evolution equations for $x$ and $e$ and expanding in these variables, we can find a
solution for the evolution of $x$ in terms of $e$ at each order as described in~\cref{sec:DCmemory}.
Here we provide $x(e)$ at 3PN and to leading order in eccentricity:
\begin{align} \label{eq:xofe3}
	x(e) = x_\newt + x_\pn1 + x_\pn{1.5} + x_\pn2 + x_\pn{2.5} + x_\pn3 \,,
\end{align}
where
\begin{subequations}
\begin{align} 
	x_\newt =&\; x_0 \left[ \left(\frac{e_0}{e}\right)^{12/19} \right]\,, \\
	x_\pn1 =&\; x_0^2 \left[ \left(\frac{e_0}{e}\right)^{24/19} \left(-\frac{2833}{3192} 
		+\frac{197 \nu}{114} \right) +\left(\frac{e_0}{e}\right)^{12/19} \left( \frac{2833}{3192} 
		-\frac{197 \nu}{114} \right) \right]\,, \\
	x_\pn{1.5} =&\; x_0^{5/2} \left[ \frac{377 \pi}{228} \left(\frac{e_0}{e}\right)^{12/19}
		-\frac{377\pi}{228} \left(\frac{e_0}{e}\right)^{30/19} \right]\,, \\
	x_\pn2 =&\; x_0^3 \biggl[ \left(\frac{e_0}{e}\right)^{12/19} \left(-\frac{358353209}{366799104} 
		+\frac{738407 \nu}{727776}+\frac{20597 \nu^2}{17328} \right) +\left( \frac{e_0}{e} 
		\right)^{24/19} \left( -\frac{8025889}{5094432} +\frac{558101 \nu}{90972} -\frac{38809 
		\nu^2}{6498}  \right) \no
		&+\left(\frac{e_0}{e} \right)^{36/19} \left( \frac{936217217}{366799104} 
		-\frac{578135\nu}{80864} +\frac{248681 \nu^2}{51984} \right) \biggr]\,, \\
	x_\pn{2.5}=&\; x_0^{7/2} \biggl[ \left(\frac{e_0}{e}\right)^{12/19} \left( -\frac{3763903 
		\pi}{7277760} -\frac{12788779 \pi \nu}{1819440} \right) +\left( \frac{e_0}{e} 
		\right)^{24/19} \left( -\frac{1068041 \pi}{363888} +\frac{74269 \pi  \nu}{12996}\right)\no
		& +\left( \frac{e_0}{e} \right)^{30/19} \left( -\frac{5340205 \pi}{1455552} + \frac{371345 
		\pi \nu}{51984} \right) +\left( \frac{e_0}{e} \right)^{42/19} \left( \frac{12956437 
		\pi}{1819440} -\frac{2651489 \pi \nu}{454860} \right) \biggr] \,,\\
	x_\pn3 =&\; x_0^4 \Biggl\{ \left(\frac{e_0}{e}\right)^{12/19} \biggl[ 
		\frac{4942027570449143}{96592876047360} +\frac{81025 \pi^2}{103968} -\frac{3317 \eg}{399} 
		-\frac{12091 \ln 2}{5985} -\frac{78003 \ln 3}{5320} -\frac{3317 \ln x_0}{798}\no
		&+\left(\frac{10309531979}{7466981760} -\frac{3977 \pi^2}{3648}\right) \nu -\frac{267351733 
		\nu^2}{82966464} -\frac{772583 \nu^3}{2222316} \biggr] +\left( \frac{e_0}{e} 
		\right)^{30/19} \left( -\frac{710645 \pi^2}{103968} \right) \no
		&+\left( \frac{e_0}{e} \right)^{24/19} \biggl( \frac{605942457431}{585411369984} 
		-\frac{3267214507\nu}{2986792704} -\frac{543796927 \nu^2}{82966464} +\frac{27463573 
		\nu^3}{2963088} \biggr) \no
		&+\left( \frac{e_0}{e} \right)^{36/19} \biggl( \frac{2652303375761}{390274246656} 
		-\frac{449767537459 \nu}{13938365952} +\frac{2754579983\nu^2}{55310976} -\frac{48990157 
		\nu^3}{1975392} \biggr) \no
		&+\left( \frac{e_0}{e} \right)^{48/19}\biggl[-\frac{1628129474693173}{27597964584960} 
		+\frac{157405\pi^2}{25992} +\frac{3317 \eg}{399} +\frac{12091 \ln 2}{5985} +\frac{78003 \ln 
		3}{5320} +\frac{6634}{2527}\ln \left(\frac{e_0}{e}\right) \no
		&+\frac{3317 \ln x_0}{798} +\left(\frac{6686551181963}{209075489280} +\frac{3977 
		\pi^2}{3648}\right) \nu -\frac{6641442629\nu^2}{165932928} +\frac{282310639 
		\nu^3}{17778528} \biggr] \Biggr\}\,.
\end{align}
\end{subequations}

\section{Oscillatory memory integral}\label{sec:oscintegral}

Here we derive the formula to evaluate the oscillatory memory integrals in~\cref{eq:oscmem}. For
convenience we set $G=c=1$ in this appendix. We define the integral that has to be computed as
\begin{align}\label{eq:Jmem}
	J_\mem &= \int_{-\infty}^{T_R}  \ud t\, x^p(t) \, e^q(t) \, \ue^{\ui (s \lambda_\xi + r \xi)}\,.
\end{align}
We follow the approach of Ref.~\cite{arun-2004}, where this integral was evaluated in the case of
circular orbits $(q=0)$. The eccentric orbit is assumed to evolve only with the secular
radiation-reaction equations given in~\cref{eq:peters-mathews} starting from $x=0$ and $e=1$ in
the remote past. Every astrophysical process like capture or mass loss possibly happening to the
binary is ignored. We start by restating the evolution equation for $x$ at leading order in $x$ and
$e$,
\begin{align}\label{eq:dxdt}
	\frac{\ud x(t)}{\ud t} = \frac{64 \nu x^5(t)}{5 m}\left[1+ \frac{157}{24} e^2(t)\right]\,,
\end{align}
and integrate it over a time interval up to some coalescence time $T_C$, where the orbital
frequency and therefore $x$ tends to infinity:
\begin{align}
	\int_{t}^{T_C} \ud t = \int_{x(t)}^{\infty} \frac{\ud x(t)}{(\ud x/\ud t)}\,.
\end{align}
Thereby, we find an explicit relation between the orbital frequency (related to $x$) and time $t$:
\begin{align}\label{eq:tofx}
	T_C - t = \frac{5m}{256 \nu}\frac{1}{x^4(t)} \left[1 -\frac{157}{43} e^2(t) \right]\,.
\end{align}
We can now invert the $x(e)$ relation derived in~\cref{eq:xofe} to find $e$ as a function of $x$.
Considering only the leading order, we find
\begin{align}\label{eq:eofx}
	e(t) = e(T_R) \left(\frac{x(T_R)}{x(t)} \right)^{19/12}\,.
\end{align}
Using~\cref{eq:tofx,eq:eofx} we get $x$ as an explicit function of $t$:
\begin{align}\label{eq:xoft}
	x(t) = \frac{1}{4}  \left(\frac{5 m}{\nu (T_C-t)}\right)^{1/4} \left[1 - \frac{157}{172}
	\,e^2(T_R) \left(\frac{T_C -t}{T_C -T_R}\right)^{19/24} \right]\,.
\end{align}
A quick check reveals that this expression indeed solves the differential equation
in~\cref{eq:dxdt}. Since the memory integral runs up to the current time $T_R$, we introduce a new
integration variable $y$ which is better suited to the integration limits we have:
\begin{align}
	y = \frac{T_R - t}{T_C - T_R}\,.
\end{align}
Next, we express the time-dependent quantities in the integral in terms of $y$ and their values at
the current time $T_R$. For $x$ we find
\begin{align}\label{eq:xy}
	x(y) = x(T_R)  (1+y)^{-1/4} \left[1 - \frac{157}{172} \,e^2(T_R) \left( (1+y)^{19/24} -1\right)
		\right]\,,
\end{align}
and for the eccentricity we find
\begin{align}\label{eq:ey}
	e(y) = e(T_R) \left(1+ y\right)^{19/48}\,.
\end{align}
Note that while going back in time, with increasing $y$, we only let the eccentricity evolve until
$e = 1$ is reached. Furthermore, we need the redefined mean anomaly $\xi(t)$ in terms of $y$ and its
value at the current time. Because $\xi$ is defined in terms of $\dot{\xi} = n$, we have to
calculate the integral
\begin{align}
	\xi(t) = \xi(T_C) - \int^t_{T_C} \ud t' \, n(t') = \xi(T_C) - \frac{1}{m} \int^t_{T_C} \ud t'
		\, x^{3/2}(t') = \xi(T_C) - \frac{(T_C - T_R)}{m} \int_{-1}^{y} \ud y' \, x^{3/2}(y') \,.
\end{align}
We can now evaluate the latter integral by inserting the expression for $x(y)$ given
in~\cref{eq:xy}. This leads to
\begin{align}
	\xi(t) = \xi(T_C) - \frac{8(T_C - T_R) x^{3/2}(T_R)}{5m}(1+y)^{5/8}\left[1 - \frac{471}{11696}
		e^2(T_R) \left(15(1+y)^{19/24} -34\right)\right] \,,
\end{align}
where $\xi(T_C)$ is the value of $\xi$ at the moment of coalescence. Thus, at the current time
$T_R$ the mean anomaly is given by
\begin{align}\label{eq:lTR}
	\xi(T_R) =&\; \xi(T_C) - \frac{8 (T_C - T_R) x^{3/2}(T_R)}{5 m} \left[1 + \frac{8949}{11696}
		e^2(T_R)\right]\,. 
\end{align}
Now we are able to express $\xi(t)$ in terms of $\xi(T_R)$ and $y$,
\begin{align}\label{eq:xit}
	\xi(t) = \xi(T_R) - \frac{8 (T_C - T_R) x^{3/2}(T_R)}{5 m}\left[(1+y)^{5/8}-1\right]\left[1 -
		\frac{471}{11696} e^2(T_R) \frac{19 -34 (1+y)^{5/8} +15(1+y)^{17/12}}{(1+y)^{5/8}
		-1}\right] \,,
\end{align}
where $x(T_R)$ and $e(T_R)$ stand for the respective current values of $x$ and $e$.

At this point, we introduce a dimensionless ``adiabatic parameter" $\chi(T_R)$, which is connected
with the inspiral rate at the current retarded time $T_R$. We define it as the ratio between the
current period and the time left until coalescence,
\begin{align}
	\chi(T_R) = \frac{1}{n(T_R)(T_C - T_R)}\,,
\end{align}
where $n(T_R) = x^{3/2}(T_R) / m$ at leading order. Explicitly in terms of $x(T_R)$ and $e(T_R)$, it
reads
\begin{align}
	\chi(T_R) = \frac{256 \nu}{5} x^{5/2}(T_R) \left[1 +\frac{157}{43}e^2(T_R)\right] \,.
\end{align}
Inserting $\chi(T_R)$ into~\cref{eq:xit}, we find
\begin{align}\label{eq:xiy}
	\xi(t) = \xi(T_R) - \frac{8}{5 \chi(T_R)} \left[(1+y)^{5/8} -1\right]\left[1 -
		\frac{471}{11696} e^2(T_R) \frac{19 -34 (1+y)^{5/8} +15(1+y)^{17/12}}{(1+y)^{5/8}
		-1}\right] \,.
\end{align}
Now we put~\cref{eq:xiy,eq:xy,eq:ey} into the oscillatory integral and write it as an integral
over $y$:
\begin{align}
	J_\mem =&\; (T_C-T_R)\int_{0}^{\infty} \ud y \, x^p(y) \, e^q(y) \, \ue^{\ui (s\lambda_\xi(y)
		+r \xi(y))}\no
		=&\;(T_C-T_R)\,  \ue^{\ui (r+s(1+k)) \xi(T_R)} \int_{0}^{\infty} \ud y \, x^p(y) \, e^q(y)
		\exp\biggl\{ -\frac{8 \ui (r+s(1+k))}{5 \chi(T_R)} \left[(1+y)^{5/8} -1\right]\no &\times
		\left[1 -\frac{471}{11696} e^2(T_R) \frac{19 -34 (1+y)^{5/8} + 15(1+y)^{17/12}}{(1+y)^{5/8}
		-1}\right] \biggr\} \,.
\end{align}
Let us look at the form of this integral:
\begin{align} \label{eq:Jstruc}
	J_\mem \sim \int_{0}^{\infty} \ud y \, f(y) \exp \left[ \frac{\ui}{\chi(T_R)}g(y) \right]\,.
\end{align}
The strategy is to integrate by parts, and therefore we need to know the following type of integral:
\begin{align}
	\int \ud y \, \ue^{\ui \sigma g(y)} = -\frac{\ui}{\sigma g'(y)} \ue^{\ui \sigma g(y)} + 
		\bigO(g'(y)^{-2}) \,.
\end{align}
This formula is valid as long as $g'(y)$ is sufficiently large. Integrating~\cref{eq:Jstruc} by
parts we get
\begin{align}\label{eq:Jpart}
	J_\mem \sim f(y) \left[ -\frac{\ui \chi(T_R)}{g'(y)} \exp \left[
		\frac{\ui}{\chi(T_R)}g(y)\right] \right]_{0}^{\infty} + \ui \chi(T_R) \int_{0}^{\infty} \ud
		y \, \frac{f'(y)}{g'(y)} \exp \left[ \frac{\ui}{\chi(T_R)}g(y) \right] \,.
\end{align}
As $y$ approaches infinity in the remote past, we notice that $f(y) = x^p(y) e^q(y)$ goes to zero. 
This is because at early times the frequency reaches zero and the eccentricity cannot grow past
$e=1$ in our model. Evaluating the first term at $y=0$, we recover $x$ and $e$ at the current time
and the exponential factor is just $1$ since $g(0) = 0$. The derivative $g'(y)$ in the denominator
evaluated at $y=0$ is effectively 1 multiplied by some constants. What remains in the first term
of~\cref{eq:Jpart} is therefore of order $\chi(T_R)$. Looking at the second term, we find the same
integral form as in~\cref{eq:Jstruc}. Successively integrating by parts would yield another factor
of $\chi(T_R)$ each time. Since this parameter is already of order 2.5PN, the higher-order
$\chi(T_R)$ contributions can be safely ignored. Including everything of order $\chi(T_R)$, we find
the formula
\begin{align}
	J_\mem &= -(T_c-T_R) \,x^p\, e^q \, \ue^{\ui (s\lambda_\xi +r \xi)} \frac{\ui \chi(T_R)}{(r
		+s(1+k))} \no
		&= - \frac{\ui}{n(r +s(1+k))} x^p\, e^q \, \ue^{\ui (s\lambda_\xi +r \xi)} \,,
\end{align}
which allows us to compute the oscillatory hereditary integrals in~\cref{sec:oscmemory}.

\section{List of DC memory modes} \label{sec:DCmemlist}

Here we list the 3PN-accurate DC memory contributions to the $h^{\ell m}$ modes at leading order in
eccentricity in the following form:
\begin{align}
	h^{\ell m}_\memdc =  \frac{8 G m \nu}{c^2 R} x \sqrt{\frac{\pi}{5}} H^{\ell m}_\memdc \,,
\end{align}
where $H^{\ell m}_\memdc$ is a function of $x$ and $e$. The nonzero modes read:
\begin{subequations}
\begin{flalign}
	H^{20}_\memdc =&\; -\frac{5}{14 \sqrt{6}} \left( H^{20}_\newt + x H^{20}_\pn{1} + x^{3/2} 
		H^{20}_\pn{1.5} +x^{2} H^{20}_\pn{2} +x^{5/2} H^{20}_\pn{2.5} +x^{3} H^{20}_\pn{3}\right) 
		\,,\\
	H^{20}_\newt =&\; 1 -\left( \frac{e}{e_i} \right)^{12/19} \,,\\
	H^{20}_\pn1 =&\;  -\frac{4075}{4032} +\frac{67 \nu}{48} +\left(\frac{e}{e_i} \right)^{12/19} 
		\left( -\frac{2833}{3192} +\frac{197 \nu}{114} \right) +\left( \frac{e}{e_i} 
		\right)^{24/19} \left( \frac{145417}{76608}-\frac{2849 \nu}{912}\right) \,,\\
	H^{20}_\pn{1.5} =&\; -\frac{377\pi}{228} \left(\frac{e}{e_i} \right)^{12/19} 
		+\frac{377\pi}{228} \left( \frac{e}{e_i}\right)^{30/19} \,,\\
	H^{20}_\pn2 =&\; -\frac{151877213}{67060224} -\frac{123815 \nu}{44352} +\frac{205 \nu^2}{352} 
		+\left(\frac{e}{e_i}\right)^{12/19} \left(\frac{358353209}{366799104} -\frac{738407 
		\nu}{727776} -\frac{20597 \nu^2}{17328}\right)\no 
		&+\left(\frac{e}{e_i}\right)^{24/19} \left(\frac{411966361}{122266368} -\frac{825950 
		\nu}{68229} +\frac{561253 \nu^2}{51984}\right) \no
		&+\left(\frac{e}{e_i}\right)^{36/19} \biggl(-\frac{50392977379}{24208740864}+ 
		\frac{764295307 \nu}{48033216} -\frac{11654209\nu^2}{1143648}\biggr) \,,\\
	H^{20}_\pn{2.5} =&\; -\frac{253 \pi}{336} +\frac{253 \pi \nu}{84} +\left( \frac{e}{e_i} 
		\right)^{12/19} \left( \frac{3763903 \pi}{7277760} +\frac{12788779 \pi \nu}{1819440}\right) 
		+\left( \frac{e}{e_i} \right)^{24/19} \left(\frac{54822209 \pi}{8733312} -\frac{1074073 \pi 
		\nu}{103968} \right) \no
		&+\left(\frac{e}{e_i}\right)^{30/19} \left(\frac{5340205 \pi}{1455552} -\frac{371345 \pi 
		\nu}{51984} \right) +\left(\frac{e}{e_i}\right)^{42/19} \left(-\frac{424020733 
		\pi}{43666560} +\frac{27049187 \pi \nu}{3638880} \right) \,,\\
	H^{20}_\pn{3} =&\;  -\frac{4397711103307}{532580106240} +\left(\frac{700464542023}{13948526592} 
		-\frac{205 \pi^2}{96}\right) \nu +\frac{69527951 \nu^2}{166053888} +\frac{1321981 
		\nu^3}{5930496} \no
		&+\left(\frac{e}{e_i}\right)^{12/19} \biggl[-\frac{4942027570449143}{96592876047360} 
		-\frac{81025 \pi^2}{103968} +\frac{3317 \eg}{399} +\left( -\frac{10309531979}{7466981760} 
		+\frac{3977 \pi^2}{3648}\right) \nu \no
		&+\frac{267351733 \nu^2}{82966464} +\frac{772583 \nu^3}{2222316} +\frac{12091 \ln 2}{5985} 
		+\frac{78003 \ln 3}{5320} +\frac{3317 \ln x}{798}\biggr] +\frac{710645 \pi^2}{103968} 
		\left( \frac{e}{e_i}\right)^{30/19} \no
		&+\left(\frac{e}{e_i}\right)^{24/19} \left(-\frac{31102835980319}{14049872879616} 
		+\frac{279737759653 \nu}{167260391424} +\frac{26730466283 \nu^2}{1991195136} 
		-\frac{397176241 \nu^3}{23704704}\right) \no
		&+\left(\frac{e}{e_i}\right)^{36/19} \left(-\frac{142763304914707}{25758100279296} 
		+\frac{48901891428821 \nu}{919932152832} -\frac{400181473249 \nu^2}{3650524416} 
		+\frac{2295879173 \nu^3}{43458624}\right) \no 
		&+\left(\frac{e}{e_i}\right)^{48/19} \biggl[\frac{385621605844415513}{5740376633671680} 
		-\frac{157405 \pi^2}{25992} -\frac{3317 \eg}{399} +\left( 
		-\frac{49590995147570629}{478364719472640} +\frac{1271 \pi^2}{1216}\right) \nu \no
		&+\frac{3194536246463\nu^2}{34514049024} -\frac{1672948713 \nu^3}{45653504}-\frac{12091 \ln 
		2}{5985} -\frac{78003 \ln 3}{5320} -\frac{3317 \ln x}{798} -\frac{6634}{2527} \ln \left( 
		\frac{e}{e_i} \right) \biggr] \,,&&
\end{flalign}
\end{subequations}
%

\begin{subequations}
\begin{flalign}
	H^{40}_\memdc =&\; -\frac{1}{504 \sqrt{2}} \left( H^{40}_\newt + x H^{40}_\pn1 + x^{3/2} 
		H^{40}_\pn{1.5} +x^{2} H^{40}_\pn2 +x^{5/2} H^{40}_\pn{2.5} +x^{3} H^{40}_\pn3 \right) \,,\\
	H^{40}_\newt =&\; 1 -\left( \frac{e}{e_i} \right)^{12/19}\,,\\
	H^{40}_\pn1 =&\; -\frac{180101}{29568} +\frac{27227 \nu}{1056} + \left( \frac{e}{e_i} 
		\right)^{12/19} \left(-\frac{2833}{3192}+\frac{197 \nu}{114} \right) + \left( \frac{e}{e_i} 
		\right)^{24/19} \biggl(\frac{3920527}{561792}-\frac{183995 \nu}{6688} \biggr) \,,\\
	H^{40}_\pn{1.5} =&\; - \frac{377 \pi}{228} \left(\frac{e}{e_i}\right)^{12/19} + \frac{377 
		\pi}{228}\left(\frac{e}{e_i}\right)^{30/19}\,,\\
	H^{40}_\pn2 =&\; \frac{2201411267}{158505984} -\frac{34829479 \nu}{432432} +\frac{844951 
		\nu^2}{27456} + \left(\frac{e}{e_i}\right)^{12/19} \left( \frac{358353209}{366799104} 
		-\frac{738407 \nu}{727776} -\frac{20597 \nu^2}{17328} \right)\no
		&+ \left(\frac{e}{e_i}\right)^{24/19} \left( \frac{11106852991}{896620032} -\frac{584029331 
		\nu}{8005536} +\frac{36247015 \nu^2}{381216} \right)\no
		&+\left( \frac{e}{e_i} \right)^{36/19} \left( -\frac{17153749047583}{629427262464} 
		+\frac{24120402175 \nu}{156107952} -\frac{1235668217 \nu^2}{9911616}\right)\,, \\
	H^{40}_\pn{2.5} =&\; -\frac{13565 \pi}{1232} +\frac{13565 \pi \nu}{308} +\left( 	
		\frac{e}{e_i}\right)^{12/19} \left( \frac{3763903 \pi}{7277760} +\frac{12788779 \pi 
		\nu}{1819440}\right) \no
		&+\left(\frac{e}{e_i}\right)^{24/19} \left( \frac{1478038679 \pi}{64044288} -\frac{69366115 
		\pi \nu}{762432} \right) +\left(\frac{e}{e_i}\right)^{30/19} \left(\frac{5340205 
		\pi}{1455552} -\frac{371345 \pi \nu}{51984}\right) \no
		&+\left(\frac{e}{e_i}\right)^{42/19} \left( -\frac{473166857 \pi}{29111040} 
		+\frac{1255597433 \pi \nu}{26685120}\right)\,,\\
	H^{40}_\pn3 =&\; \frac{15240463356751}{781117489152} +\left( -\frac{1029744557245}{27897053184} 
		-\frac{205 \pi^2}{96} \right) \nu -\frac{4174614175 \nu ^2}{36900864} +\frac{221405645 
		\nu^3}{11860992} \no
		&+\left(\frac{e}{e_i}\right)^{12/19} \biggl[ -\frac{4942027570449143}{96592876047360} 
		-\frac{81025 \pi^2}{103968} +\frac{3317 \eg}{399} +\left( 
		-\frac{10309531979}{7466981760} +\frac{3977 \pi^2}{3648} \right) \nu \no
		&+\frac{267351733 \nu^2}{82966464} +\frac{772583 \nu^3}{2222316} +\frac{12091 \ln 2}{5985} 
		+\frac{78003 \ln 3}{5320} +\frac{3317 \ln x}{798} \biggr] +\frac{710645 \pi^2}{103968} 
		\left( \frac{e}{e_i} \right)^{30/19} \no
		&+\left( \frac{e}{e_i} \right)^{24/19} \left( -\frac{838550569998089}{103032401117184} 
		+\frac{30467243664175 \nu}{1226576203776} +\frac{963631094693 \nu^2}{14602097664} 
		-\frac{25650558955 \nu^3}{173834496}\right) \no 
		&+\left( \frac{e}{e_i} \right)^{36/19} \left( -\frac{48596571051802639}{669710607261696} 
		+\frac{13219254870469451 \nu}{23918235973632} -\frac{107533340184449 \nu^2}{94913634816} 
		+\frac{243426638749 \nu^3}{376641408} \right) \no
		&+\left( \frac{e}{e_i} \right)^{48/19} \biggl[ \frac{1289915690995598063}{11480753267343360}
		-\frac{157405 \pi^2}{25992} -\frac{3317 \eg}{399} +\left( 
		-\frac{515898615572711953}{956729438945280} +\frac{1271 \pi^2}{1216}\right) \nu \no
		&+\frac{297870712456705 \nu^2}{253103026176} -\frac{520032054523 \nu^3}{1004377088} 
		-\frac{12091 \ln 2}{5985} -\frac{78003 \ln 3}{5320} -\frac{3317 \ln x}{798} 
		-\frac{6634}{2527} \ln \left(\frac{e}{e_i}\right)\biggr] \,,&& 
\end{flalign}
\end{subequations}
%

\begin{subequations}
\begin{flalign}
	H^{60}_\memdc =&\; \frac{4195}{1419264 \sqrt{273}} \left(x H^{60}_\pn1 +x^{2} 
		H^{60}_\pn2 +x^{5/2} H^{60}_\pn{2.5} +x^{3} H^{60}_\pn3 \right) \,,\\
	H^{60}_\pn1 =&\; 1 -\frac{3612 \nu}{839} \,,\\
	H^{60}_\pn2 =&\; -\frac{45661561}{6342840} +\frac{101414 \nu}{2517} -\frac{48118 
		\nu^2}{839} +\left(\frac{e}{e_i}\right)^{24/19} \left(-\frac{2833}{1596} +\frac{530740 
		\nu}{47823} -\frac{237188 \nu^2}{15941}\right) \no
		&+\left(\frac{e}{e_i}\right)^{36/19} \left( \frac{1081489489}{120513960} -\frac{819202 
		\nu}{15941} +\frac{1151430 \nu^2}{15941} \right) \,,\\
	H^{60}_\pn{2.5} =&\; \frac{1248 \pi}{839} -\frac{4992 \pi \nu}{839} +\left( \frac{e}{e_i} 
		\right)^{24/19} \biggl(-\frac{377 \pi}{114} +\frac{226954 \pi \nu}{15941} \biggr) 
		+\left(\frac{e}{e_i}\right)^{42/19} \biggl(\frac{174031 \pi}{95646} -\frac{132106 \pi 
		\nu}{15941} \biggr)\,,\\
	H^{60}_\pn3 =&\; \frac{3012132889099}{144921208320} -\frac{27653500031 \nu}{191694720} 
		+\frac{1317967427 \nu^2}{4107744} -\frac{24793657 \nu^3}{342312} \no
		&+\left(\frac{e}{e_i}\right)^{24/19} \biggl(\frac{213887207}{183399552} -\frac{7295329871 
		\nu}{1831812192} -\frac{214435261 \nu^2}{21807288} +\frac{41962109 \nu^3}{1817274} \biggr) 
		\no
		&+\left( \frac{e}{e_i}\right)^{36/19} \biggl(\frac{3063859722337}{128226853440} 
		-\frac{839669231153 \nu}{4579530480} +\frac{555765673 \nu^2}{1211516} -\frac{113415855 
		\nu^3}{302879}\biggr) \no
		&+\left(\frac{e}{e_i}\right)^{48/19} \biggl(-\frac{9789584507539}{213536964096} 
		+\frac{206521649193667 \nu}{622816145280} -\frac{380487275717 \nu^2}{494298528} 
		+\frac{17456918535 \nu^3}{41191544} \biggr)\,, &&
\end{flalign}
\end{subequations}
%
\begin{subequations}
\begin{flalign}
	H^{80}_\memdc =&\; -\frac{75601}{213497856 \sqrt{119}} \left(x^{2} H^{80}_\pn2 +x^{3} 
		H^{80}_\pn3\right) \,,\\
	H^{80}_\pn2 =&\; 1-\frac{452070 \nu}{75601} +\frac{733320 \nu^2}{75601} +\left( \frac{e}{e_i} 
		\right)^{36/19} \left( -1 +\frac{452070 \nu}{75601} -\frac{733320 \nu^2}{75601} \right)\,,\\
	H^{80}_\pn3 =&\; -\frac{265361599}{33869248} +\frac{18177898147 \nu}{321757856} 
		-\frac{722521125 \nu^2}{5745676} +\frac{261283995 \nu^3}{2872838} \no
		&+\left(\frac{e}{e_i}\right)^{36/19} \biggl(-\frac{2833}{1064}+\frac{848864713 
		\nu}{40219732} -\frac{81627030 \nu^2}{1436419} +\frac{72232020 \nu^3}{1436419} \biggr)\no
		&+\left(\frac{e}{e_i}\right)^{48/19} \biggl(\frac{50791665}{4838464} -\frac{3566973693 
		\nu}{45965408} +\frac{1049029245 \nu^2}{5745676} -\frac{405748035 \nu^3}{2872838} 
		\biggr)\,,&&
\end{flalign}
\end{subequations}
%

\begin{subequations}
\begin{flalign}
	H^{10\,0}_\memdc =&\; \frac{525221}{6452379648 \sqrt{154}} x^3 \biggl[ 1-\frac{79841784 
		\nu}{9979199} +\frac{198570240 \nu^2}{9979199} -\frac{172307520 \nu^3}{9979199} \no
		&+\left(\frac{e}{e_i}\right)^{48/19} \biggl(-1+\frac{79841784 \nu}{9979199} 
		-\frac{198570240 \nu^2}{9979199} +\frac{172307520 \nu^3}{9979199} \biggr) \biggr]\,.&&
\end{flalign}
\end{subequations}

\section{List of oscillatory memory modes} \label{sec:oscmemlist}

Here we list the nonzero oscillatory memory contributions to the $h^{\ell m}$ modes at 3PN order 
and to quadratic order in eccentricity in the following way:
\begin{align}
	h^{\ell m}_\memosc = \frac{8 G m \nu}{c^2 R} x \sqrt{\frac{\pi}{5}}  \ue^{-\ui m \psi} H^{\ell 
		m}_\memosc \,,
\end{align}
where $H^{\ell m}_\memosc$ is a function of $x$, $e$, and the modified mean anomaly $\xi$. To 
improve readability in the odd-$m$ expressions, we define $\Delta = (m_1 - m_2)/m = \sqrt{1-4\nu}$:
\begin{subequations}
\begin{align}
	H^{20}_\memosc =&\; \frac{ \sqrt{6}\, \ui \,x^{5/2} e  \nu}{7} \biggl[-16 \ue^{-\ui \xi} +16 
		\ue^{\ui \xi} -\frac{647}{36} e \ue^{-2 \ui \xi} +\frac{647}{36} e \ue^{2 \ui \xi} 
		\biggr]\,,\\
	H^{22}_\memosc =&\; \ui \, x^{3/2} \, e^2 \nu \, \ue^{2\ui \xi} \left[ -\frac{13}{252} + 
		\left(\frac{697}{336} -\frac{865 \nu}{216} \right)x -\frac{29\pi}{126} x^{3/2}\right] \no
		&+  \frac{\ui x^{5/2} e \nu}{21} \left[ \frac{19}{6}e +\frac{4}{3} \ue^{-\ui \xi} - 4 
		\ue^{\ui \xi} +\frac{65}{24} e \ue^{-2\ui \xi}\right] \,,\\
	H^{31}_\memosc =&\; \frac{\sqrt{14} x^2 \nu \Delta}{90}\biggl\{ \frac{44}{3} e^2 -\frac{44}{3} 
		e\, \ue^{\ui \xi} -\frac{44}{3} e^2 \ue^{2 \ui \xi} +x \biggl[-\frac{121}{7} -\frac{43}{2} 
		e \ue^{-\ui \xi } -\frac{2987}{84} e^2 \ue^{-2 \ui \xi}\no
		&+e \ue^{\ui \xi} \left(\frac{19801}{264} -\frac{2521 \nu}{231}\right) +e^2 \ue^{2 \ui \xi} 
		\left(\frac{7957}{88} -\frac{827 \nu}{231}\right) +e^2 \left(- \frac{111821}{616} + 
		\frac{827 \nu}{231} \right)\biggl]\biggr\}\,,\\
	H^{33}_\memosc =&\; \frac{x^3 \nu \Delta}{6 \sqrt{210}} \left[ \frac{22}{9} +19 e^2 +11 e\, 
		\ue^{-\ui \xi} +\frac{1}{3} e\, \ue^{\ui \xi}+\frac{713}{30} e^2 \ue^{-2 \ui \xi} 
		-\frac{119}{6} e^2 \ue^{2 \ui \xi}\right]\,,\\
	H^{40}_\memosc =&\; \frac{\sqrt{2} \ui \,x^{5/2} e  \nu}{210} \biggl[-8 \ue^{-\ui \xi} +8 
		\ue^{\ui \xi} -\frac{143}{16} e\, \ue^{-2 \ui \xi} +\frac{143}{16} e \,\ue^{2 \ui 
		\xi}\biggr]\,,\\
	H^{42}_\memosc =&\; \frac{\ui\,x^{3/2} e^2 \nu \ue^{2\ui \xi}}{216 \sqrt{5}} 
		\left[-\frac{13}{14} +x \left(\frac{7943}{56} -\frac{25393 \nu}{66} \right) -\frac{29}{7} 
		\pi x^{3/2}\right]\no
		&+\frac{\ui\, x^{5/2} e \nu}{126 \sqrt{5}} \left[\frac{19}{12} e +\frac{2}{3} \ue^{-\ui 
		\xi} -2 \ue^{\ui \xi} +\frac{65}{48} e \,\ue^{-2 \ui \xi}\right] \,,\\
	H^{44}_\memosc =&\; \frac{\ui\, x^{5/2} \nu}{6 \sqrt{35}} \biggl[\frac{2}{3} +\frac{331 
		e^2}{240} + \frac{14}{15} e\, \ue^{-\ui \xi} +2 e\, \ue^{\ui \xi} +\frac{1037}{720} e^2 
		\ue^{-2 \ui \xi} +\frac{217}{48} e^2 \ue^{2 \ui \xi} \biggr]\,,\\
	H^{51}_\memosc =&\; \frac{x^2 \nu \Delta}{18 \sqrt{385}} \biggl\{ \frac{43 e^2}{12} - 
		\frac{43}{12} e \,\ue^{\ui \xi} -\frac{43}{12} e^2 \ue^{2 \ui \xi} +x \biggl[ -\frac{26}{7} 
		-\frac{251}{56} e \,\ue^{-\ui \xi} -\frac{1199}{168} e^2 \ue^{-2 \ui \xi} \no
		&+ e\, \ue^{\ui \xi} \left(\frac{8627}{156} -\frac{41807 \nu}{312}\right) +e^2 \ue^{2 \ui 
		\xi} \left(\frac{785}{13} -\frac{5156 \nu}{39}\right) +e^2 \left(-\frac{8321}{104} 
		+\frac{5156 \nu}{39}\right) \biggr] \biggr\} \,,\\
	H^{53}_\memosc =&\; \frac{x^3 \nu \Delta}{2\sqrt{330}} \biggl[-\frac{2}{189} +\frac{27 
		e^2}{112} + \frac{41}{336} e\, \ue^{-\ui \xi} -\frac{67}{504} e \ue^{\ui \xi} +\frac{1531
		e^2 \ue^{-2 \ui \xi}}{5040} -\frac{47}{72} e^2 \ue^{2 \ui \xi} \biggr]\,,\\
	H^{55}_\memosc =&\; \frac{x^3 \nu \Delta}{14 \sqrt{66}} \biggl[\frac{18}{5} +\frac{8909
		e^2}{720} + \frac{457}{72} e\, \ue^{-\ui \xi} +\frac{197}{16} e\, \ue^{\ui \xi}
		+\frac{787}{72} e^2 \ue^{-2 \ui \xi} +\frac{4369}{144} e^2 \ue^{2 \ui \xi}\biggr]\,,\\
	H^{62}_\memosc =&\; \frac{\ui\,x^{5/2} e^2 \nu \, \ue^{2 \ui \xi}}{352 \sqrt{65}}
		\biggl[\frac{2783}{168} - 53 \nu \biggr]\,,\\
	H^{71}_\memosc =&\; \frac{5x^3 \nu e \Delta}{30888 \sqrt{2}} \biggl[ \ue^{\ui \xi} \left(
		\frac{5023}{168} - 97 \nu \right) +e\, \ue^{2 \ui \xi } \left(\frac{5023}{168} -97 \nu
		\right) +e \left(-\frac{5023}{168} +97 \nu \right)\biggr] \,.
\end{align}
\end{subequations}

\end{widetext}

\bibliographystyle{apsrev4-1}
\bibliography{references-mem}

\end{document}